\documentclass[twocolumn,showpacs,preprintnumbers,amsmath,amssymb]{revtex4}

\usepackage{graphicx}
\usepackage{dcolumn}
\usepackage{bm}

\begin{document}

\preprint{APS/123-QED}

\title{ Neutrino mass effects on vector and tensor  CMB anisotropies in the presence of a primordial magnetic field}

\author{Kazuhiko Kojima}
\email{kojima@th.nao.ac.jp}
\affiliation{%
Department of Astronomy, University of Tokyo,\\
7-3-1 Hongo, Bunkyo-ku, Tokyo 113-0033, Japan
}%
\affiliation{Division of Theoretical Astrophysics, National Astronomical Observatory, \\
2-21-1 Osawa, Mitaka, Tokyo 181-8588, Japan
}%
\author{Kiyotomo Ichiki}
\affiliation{Research Center for the Early Universe,
School of Science, the University of Tokyo,\\
7-3-1 Hongo, Bunkyo-ku 113-0033, Japan
}%
\author{Dai G. Yamazaki}
\affiliation{Division of Theoretical Astrophysics, National Astronomical Observatory, \\
2-21-1 Osawa, Mitaka, Tokyo 181-8588, Japan
}%
\author{Toshitaka Kajino}%
\affiliation{Department of Astronomy, University of Tokyo,\\
7-3-1 Hongo, Bunkyo-ku, Tokyo 113-0033, Japan}
\affiliation{Division of Theoretical Astrophysics, National Astronomical Observatory, \\
2-21-1 Osawa, Mitaka, Tokyo 181-8588, Japan
}%
\author{Grant J. Mathews}
\affiliation{Center for Astrophysics, Department of Physics, University of Notre Dame, \\
Notre Dame, IN 46556, U.S.A.}
\affiliation{Division of Theoretical Astrophysics, National Astronomical Observatory,\\ 
2-21-1 Osawa, Mitaka, Tokyo 181-8588, Japan}

\date{\today}

\begin{abstract}
If a primordial magnetic field (PMF) is present during photon decoupling and afterward, a finite neutrino mass can affect all  modes of the CMB.  In this work, we expand on  earlier studies of the scalar mode effects by constructing  the vector and tensor mode equations in the presence of  massive neutrinos and a PMF. We compute the power spectrum of the various modes in an illustrative example and  find that the neutrino mass can significantly affect the vector and tensor modes  when a PMF exists, while the effects are negligible for no PMF. The most prominent result of the present analysis is the behavior of the EE component of the tensor mode at low multipoles.  For massive neutrinos the EE mode  can become comparable to the observed primary anisotropy.  Therefore, if and when the EE mode power spectrum is measured at low multipoles  the possibility exists  to place a strong  constraint on the sum of the neutrino masses. 

\end{abstract}

\maketitle

\section{Introduction}

The possibility that a primordial magnetic field (PMF) could affect the CMB power spectrum is a subject of considerable recent interest, e.g.~ \cite{2006ApJ...646..719Y,2004PhRvD..70d3518L,2004PhRvD..70d3011L,2002PhRvD..65l3004M,2007PhRvD..75b3002K}. Such effects arise in two ways.  Baryons are affected by the Lorentz force from a PMF.  They then  influence  the CMB spectrum indirectly through Thomson scattering.  Moreover, the Lorentz force from a PMF also has  a significant  effect on the later development of large-scale structure.   Among other things a PMF could explain  the possible power excess observed by ACBAR \cite{2004ApJ...600...32K,2008arXiv0801.1491R} for large multipoles ($l>2500$) in the CMB power spectrum. It could also explain the BB mode CMB anisotropies observed by CBI \cite{2003ApJ...591..540M} as well as  the origin of the  magnetic field inferred from the observed 
\cite{2006ApJ...637...19X,2001ApJ...547L.111C,1992ApJ...388...17W,1992ApJ...387..528K} polarization of light from  galactic  clusters. Such studies  suggest that a PMF may indeed exist, and that it is worthwhile to examine other possible consequences of its existence. In previous work \cite{kojima}, we deduced  neutrino mass effects from the  scalar mode of the CMB power spectrum using the WMAP-3yr data \cite{2007ApJS..170..377S}, when the PMF effect is also taken into consideration. Indeed, the neutrino mass constraint from the scalar mode of the CMB is the recent focus in the literature \cite{2007ApJS..170..377S,2005PhRvD..71d3001I,2006PhRvD..74b7302F}. In this work, we expand on this discussion by introducing new studies into the neutrino-mass effects  on the vector and tensor modes of the CMB as well as the scalar mode.

The CMB power spectrum induced by the PMF separates into 3 parts, i.e.~the  scalar, vector, and tensor modes. The scalar mode affects the power spectrum at  large angular scales $l<10$, while the vector mode primarily affects $l>1000$.  
In previous work \cite{kojima}, we deduced  neutrino mass constraints from the  scalar mode of the CMB power spectrum using the WMAP-3yr data.  We showed that a  PMF can decrease the upper limit on the neutrino mass. 
Our ultimate  goal, however,  is to constrain the neutrino mass using all PMF modes and using all available observations including all multipoles. However, before calculating
the neutrino mass constraint, we need to derive a framework in which to  study the effects of the neutrino mass on the CMB power spectrum in the presence of a  PMF.  Hence, 
the purpose of  this work, is to construct the relevant equations for the effects of massive neutrinos on the  vector, and tensor modes in addition to the scalar modes of the CMB power spectrum.   With this framework in place, we then study the possible neutrino mass  effects in the context of an illustrative model.  We show that  the  neutrino mass can cause the EE mode at low multipole  to become comparable to the observed primary anisotropy. We also find that the BB mode is affected by the neutrino mass and may be detectable  if the gravitational wave contribution is not too large.
Thus, if and when the EE and BB mode power spectra are measured  the possibility exists  to better constrain  the sum of the neutrino masses. 

\section{Equations}
Here, we derive equations for the perturbations on the CMB anisotropies from the various modes in the presence of a PMF.   Usually, only the scalar mode CMB anisotropies have been calculated as  in 
Ref.~ \cite{1995ApJ...455....7M}.  However, in the presence of a PMF  the  vector and tensor modes \cite{2007PhRvD..75b3002K}, are also important. Our equations are based upon the formulation given in Refs.~\cite{2008arXiv0801.2572Y,1997PhRvD..56..596H}. The equations in 
Ref.~\cite{2008arXiv0801.2572Y}
are almost the same as in Ref.~\cite{1997PhRvD..56..596H} except for the PMF terms. However, a finite neutrino mass was not considered in Ref.~\cite{2008arXiv0801.2572Y}. Hence, in this paper we  extend the derivation  in Ref.~\cite{1997PhRvD..56..596H} to deduce equations which include the effects of a finite neutrino mass.
\subsection{Boltzmann equations and anisotropies}
To begin, we choose metric perturbations in terms of the scale factor $a$ and the conformal Newtonian-gauge metric parameters $\Psi$ and $\Phi$ of the form, 
\begin{eqnarray}
\delta g_{00}=-2a^2\Psi Q^{(0)},^{~~}\delta g_{ij}=2a^2\Phi Q^{(0)}\delta_{ij}~~,
\end{eqnarray}
for the scalar mode, and
\begin{eqnarray}
\delta g_{ij}=2a^2H_T^{(1)}Q^{(1)}_{ij}~~,
\end{eqnarray}
for the vector mode, while the tenor mode metric perturbation is given by
\begin{eqnarray}
\delta g_{ij}=2a^2H^{(2)}_TQ^{(2)}_{ij}~~.
\end{eqnarray}
Here, $Q^{(0)}$, $Q^{(1)}_{ij}$ and $Q^{(2)}_{ij}$ are the harmonic modes for the scalar, vector and tensor components, respectively, while the $H_T^{(1)}, H_T^{(2)}$ are vector and tensor metric perturbations, respectively.  Here and throughout we use  the same notation as in 
Ref.~\cite{1997PhRvD..56..596H}. Our gauge for the  vector mode, however,  is different from that of Ref.~\cite{1997PhRvD..56..596H}, and the same as in Refs.  \cite{2004PhRvD..70d3518L,2004PhRvD..70d3011L}, i.e.~the perturbations for the vector mode are in space-space components rather than space-time components. This is because it is difficult to treat massive neutrinos in the gauge of  Ref.~\cite{1997PhRvD..56..596H}. 
 
 Next, we define coordinates  by setting $\hat{k}=\hat{e}_3$, and defining basis vectors: 
\begin{eqnarray}
\hat{e}^{(\pm)}=\frac{-i}{\sqrt[]{\mathstrut 2}}(\hat{e}_1\pm i\hat{e}_2),~~e^{(\pm\pm)}_{ij}=\sqrt[]{\mathstrut \frac{3}{2}}\hat{e}^{(\pm)}_i\otimes\hat{e}^{(\pm)}_j~.
\end{eqnarray}
 We also write  the distribution functions for massive neutrinos as 
 \begin{equation}
 f(q,n^i,k^i,\eta)=\bar{f}(q)+\delta f(q,n^i,k^i,\eta)\equiv \bar{f}(1+\Theta_h)~~,
 \end{equation}
  where $\delta f$ is the perturbed part, $q$ is the comoving momentum, $n^i$ is its direction, $\eta$ is the conformal time, and $\Theta_h$ is the normalized perturbation, with the subscript $h$ denoting hot dark matter. Hereafter, we attach subscript $h$ to distinguish any quantities $X_h$ which are associated with hot dark matter (i.e. massive neutrinos).  We will confine the subscript $\nu$ to denote for massless neutrinos, i.e.~$X_{\nu}$.
 
 Neutrinos do not interact with other particles.  Hence, they obey the collisionless Boltzmann equation.
 We use the  linearized collisionless Boltzmann equation to calculate the evolution of the perturbation, i.e.
 \begin{eqnarray}
\dot{\delta f}&=&- \frac{iqkn^i\hat{k}_i}{\epsilon}\delta f+(q\dot{\Phi}+ik\epsilon n^i\hat{k}_i \Psi)\frac{\partial \bar{f}}{\partial q}\nonumber\\
&=& -\frac{iqk}{\epsilon}\sqrt[]{\mathstrut \frac{4\pi}{3}}Y^{0}_1\delta f+(q\dot{\Phi}+ik\epsilon\sqrt[]{\mathstrut \frac{4\pi}{3}}Y^{0}_1\Psi)\frac{\partial \bar{f}}{\partial q}~~,
\end{eqnarray}
for the scalar mode,  while the vector mode equation is \cite{1986PhRvD..33.1576K},
\begin{eqnarray}
\dot{\delta f}&=& -\frac{iqkn^i\hat{k}_i}{\epsilon}\delta f-iqn^i\hat{k}_in^j\hat{e}^{(+)}_{j}\dot{H}^{(1)}_T\frac{\partial \bar{f}}{\partial q}\nonumber\\
&=&-\frac{iqk}{\epsilon}\sqrt[]{\mathstrut \frac{4\pi}{3}}Y^{0}_1\delta f+q\dot{H}_T^{(1)}\sqrt[]{\mathstrut \frac{4\pi}{15}}Y^{1}_2\frac{\partial \bar{f}}{\partial q}~~,
\end{eqnarray}
and the tensor mode equation is
\begin{eqnarray}
\dot{\delta f} &=&-\frac{iqkn^i\hat{k}_i}{\epsilon}\delta f+qe^{(++)}_{ij}n^in^j\dot{H}_T^{(2)}\frac{\partial \bar{f}}{\partial q}\nonumber\\
&=&-\frac{iqk}{\epsilon}\sqrt[]{\mathstrut \frac{4\pi}{3}}Y^{0}_1\delta f-q\dot{H}_T^{(2)}\sqrt[]{\mathstrut \frac{4\pi}{5}}Y^{2}_2\frac{\partial \bar{f}}{\partial q}~~,
\end{eqnarray}
where $\epsilon=\sqrt{\mathstrut q^2 +m_{\nu}^2a^2}$, in the same notation as in Ref.~\cite{1995ApJ...455....7M}.  To get hierarchial equations similar to Eq. (60) of Ref.~\cite{1997PhRvD..56..596H}, we next expand the  perturbation in spherical harmonics,
\begin{eqnarray}
\Theta_h=\sum_{l,m} (-i)^{l}\sqrt[]{\mathstrut 4\pi/(2l+1)}Y_l^m\Theta_{hl}^{(m)}  ~~,
\end{eqnarray}
where the $\Theta_{hl}^{(m)}$ are amplitudes for the various modes.  We  use the Clebsh-Goldan relation:
\begin{eqnarray}
&&{\sqrt[]{\mathstrut \frac{4\pi}{3}}}Y_1^0Y_l^m =\frac{\sqrt[]{\mathstrut l^2-m^2}}
{\sqrt[]{\mathstrut (2l+1)(2l-1)}}Y_{l-1}^m \nonumber\\
&&^{~~~~~~~~~~~~~~~~~}+\frac{\sqrt[]{\mathstrut (l+1)^2-m^2}}
{\sqrt[]{\mathstrut(2l+1)(2l+3)}}Y_{l+1}^m~~,
\end{eqnarray}
to expand the Boltzmann equations in terms of the $\Theta_{hl}^{(m)}$:
\begin{eqnarray}
&&\dot{\Theta}_{hl}^{(m)}=\frac{qk}{\epsilon}\Bigl( \Bigr. \frac{\sqrt[]{\mathstrut l^2-m^2}}{2l+1}\Theta_{hl-1}^{(m)}\nonumber\\
&&~~~~~~-\frac{\sqrt[]{\mathstrut (l+1)^2-m^2}}{2l+3}\Theta_{hl+1}^{(m)}\Bigl. \Bigr)+S_{hl}^{(m)}~~,
\label{eq101}
\end{eqnarray}
where, the source terms $S_{hl}^{(m)}$ are given by,
\begin{eqnarray}
&&S_{h0}^{(0)}=\dot{\Phi}\frac{\partial \ln \bar{f}}{\partial \ln q},^{~~}S_{h1}^{(0)}=-\frac{k\epsilon}{q}\Psi\frac{\partial \ln \bar{f}}{\partial \ln q}~~,\nonumber\\
&&S_{h2}^{(1)}=\frac{\dot{H}_T^{(1)}}{\sqrt[]{\mathstrut 3}}\frac{\partial \ln \bar{f}}{\partial \ln q},{~~}S_{h2}^{(2)}=\dot{H}_T^{(2)}\frac{\partial \ln \bar{f}}{\partial \ln q}~~.
\end{eqnarray}
This is very similar to Eq. (60) of Ref.~\cite{1997PhRvD..56..596H}, except for the source terms. This difference is due to the finite neutrino mass which makes it impossible to integrate the distribution functions easily. 
The $\Theta_{hl}^{(m)}$ are related to the density perturbation $\delta_{h}$, velocity $v_{h}$, and anisotropic stress $\pi_h$ as follows:
\begin{eqnarray}
&&\delta_h^{(0)}=a^{-4}\frac{4\pi}{\rho_h}\int q^2dq\epsilon\bar{f}\Theta_{h0}^{(0)}~~,\nonumber\\
&&v_h^{(0)}=a^{-4}\frac{4\pi/3}{\rho_h+p_h}\int q^2dqq\bar{f}\Theta_{h1}^{(0)}~~,\nonumber\\
&&\pi_h^{(0)}=a^{-4}\frac{4\pi/5}{p_h}\int q^2dq\frac{q^2}{\epsilon}\bar{f}\Theta_{h2}^{(0)}~~,\nonumber\\
&&v_h^{(1)}=a^{-4}\frac{4\pi/3}{\rho_h+p_h}\int q^2dqq\bar{f}\Theta_{h1}^{(1)}~~,\nonumber\\
&&\pi_h^{(1)}=a^{-4}\frac{8{\sqrt[]{\mathstrut 3}}\pi/15}{p_h}\int q^2dq\frac{q^2}{\epsilon}\bar{f}\Theta_{h2}^{(1)}~~,\nonumber\\
&&\pi_h^{(2)}=a^{-4}\frac{8\pi/15}{p_h}\int q^2dq\frac{q^2}{\epsilon}\bar{f}\Theta_{h2}^{(2)}~~,
\label{eq100}
\end{eqnarray}
where $\rho_h$ and $p_h$ are the energy densities and pressures from hot dark matter, i.e. massive neutrinos.
Equations for the other components are the same as in Ref.~\cite{1997PhRvD..56..596H} except for the gauge difference in the vector mode. This gauge difference changes Eq. (61) of Ref.~\cite{1997PhRvD..56..596H} to 
\begin{eqnarray}
&&S_1^{(1)}=\dot{\tau}v_B^{(1)},^{~~}S_2^{(1)}=\dot{\tau}P^{(1)}-\frac{1}{\sqrt[]{\mathstrut 3}}\dot{H}_T^{(1)}~~.
\end{eqnarray}
In order to obtain the power spectrum of the CMB and its polarization, we need to expand the temperature perturbation $\Theta$ with 
spherical harmonics, and expand the polarization fluctuation $Q\pm iU$ with spin-2 harmonics \cite{1997PhRvD..56..596H}. The expansion coefficients are $\Theta_l^{(m)}$, and $E_l^{(m)}\pm iB_l^{(m)}$ respectively, and their integral solutions are given as follows \cite{1997PhRvD..56..596H},  
\begin{eqnarray}
&&\frac{\Theta_{l}^{(0)}(\eta_0,k)}{2l+1}=\int^{\eta_0}_{0}e^{-\tau}\left[\right.(\dot{\tau}\Theta_0^{(0)}+\dot{\tau}\Psi+\dot{\Psi}-\dot{\Phi})j^{(00)}_l\nonumber\\
&&^{~~~~~~~~~~~~~}+\dot{\tau}v_B^{(0)}j_l^{(10)}+\dot{\tau}P^{(0)}j_l^{(20)}\left.\right]
\label{eq102}\\
&&\frac{\Theta_{l}^{(1)}(\eta_0,k)}{2l+1}= \int_0^{\eta_0}d\eta e^{-\tau}\Bigl[\Bigr.\dot{\tau}v_B^{(1)}j_l^{(11)}\nonumber\\
&&^{~~~~~~~~~~~~~}+\left(\dot{\tau}P^{(1)}-\dot{H}_T^{(1)}/\sqrt[]{\mathstrut 3}\right)j_l^{(21)}\Bigl.\Bigr]
\label{eq103}\\
&&\frac{\Theta_{l}^{(2)}(\eta_0,k)}{2l+1}= \int_0^{\eta_0}d\eta e^{-\tau}[\dot{\tau}P^{(2)}-\dot{H}_T^{(2)}]j_l^{(22)}
\label{eq104}\\
&&\frac{E_{l}^{(m)}(\eta_0,k)}{2l+1}=-\sqrt[]{\mathstrut 6}\int_0^{\eta_0}d\eta \dot{\tau}e^{-\tau} P^{(m)}\epsilon_{l}^{(m)}
\label{eq105}\\
&&\frac{B_{l}^{(m)}(\eta_0,k)}{2l+1}=-\sqrt[]{\mathstrut 6}\int_0^{\eta_0}d\eta \dot{\tau}e^{-\tau} P^{(m)}\beta_{l}^{(m)},
\label{eq106}
\end{eqnarray}
where $\eta_0$ is the present conformal time, while the radial temperature function $j_l^{(l^{\prime}m)}(x)$, the radial E function $\epsilon_l^{(m)}(x)$ and the radial B function $\beta_l^{(m)}(x)$ are evaluated at $x=k(\eta_0-\eta)$. Here, we have used the anisotropic scattering source $P^{(m)}\equiv (\Theta_2^{(m)}-\sqrt[]{\mathstrut 6}E_2^{(m)})/10$. The equation for the vector anisotropies is different from that of Ref.~\cite{1997PhRvD..56..596H} because we are in a different gauge. 

From Eqs.~(\ref{eq102}) - (\ref{eq106}), we can then derive  the CMB power spectra of temperature and polarization anisotropies by constructing  correlation function \cite{1997PhRvD..56..596H};
\begin{eqnarray}
&&(2l+1)^2C_l^{X\tilde{X}(m)}\nonumber\\
&&~~~~=\frac{2}{\pi}\int \frac{dk}{k} k^3X_l^{(m)\ast}(\eta_0,k)\tilde{X}_l^{(m)}(\eta_0,k)~~,
\label{eq115}
\end{eqnarray}
where $X$ is $\Theta$, $E$ or $B$.

The evolution of the perturbation variables is given by  the Einstein equations \cite{1997PhRvD..56..596H} to be:
\begin{eqnarray}
&&k^2\Phi=4\pi Ga^2\Bigl[(\rho_f\delta_f+\rho_{\gamma}\rho_{PMF}^{~})\nonumber\\
&&~~~~~~~~~~~~~~~+3\frac{\dot{a}}{a}(\rho_f+p_f)v_f^{(0)}/k\Bigr]~~,\nonumber\\
&&k^2(\Psi+\Phi)=-8\pi Ga^2(p_f\pi^{(0)}_f+p_{\gamma}\pi_{PMF}^{(0)})~~,
\label{eq107}
\end{eqnarray}
for the scalar mode, where $\rho_f$ and $p_f$ are the energy densities and pressures from the nonrelativistic (e.g.~matter) and relativistic (e.g.~photon)  fluids (following the notation in Ref.  \cite{1997PhRvD..56..596H}). $\rho_{PMF}$ and $\pi_{PMF}^{(m)}$ are the energy density and anisotropic stress from the PMF, which are normalized by the photon density $\rho_{\gamma}$ and photon pressure $p_{\gamma}$, respectively, for the following reason.
Energy momentun tensor for the magnetic field is defined by:
\begin{eqnarray}
T^{\rm PMF}_{00}&=& \frac{1}{8\pi} B^2 
\label{eq125} \\
T^{\rm PMF}_{ij}&=& \frac{1}{4\pi}\bigl( B_iB_j-\frac{1}{2}\delta_{ij}B^2\bigr)~~,
\label{eq126}
\end{eqnarray} 
where we have neglected the electric field because the coductance of the early universe is assumed to be very large
by taking the MHD approximation.
Transforming Eqs. (\ref{eq125}) and (\ref{eq126}) into wavenumber-space and decomposing the traceless part of Eq. (\ref{eq126}) into scalar, vector and tensor mode, we can define the energy density and anisotropic stress for the PMF.
 Since we assume that the conductance is
infinite, the magnetic field is "frozen in" and the time evolution of $B$ is $B\propto a^{-2}$. 
This means that the energy density and anisotropic stress for the PMF grow as $\propto a^{-4}$. 
To eliminate the effect of expansion of the universe, we define
 the energy density and anisotropic stress for the PMF divided $\rho_{\gamma}~(\propto a^{-4})$ and  $p_{\gamma}~(\propto a^{-4})$, respectively, as shown in Eq. (\ref{eq107}). In our definition, therefore,
 $\rho_{PMF}$ and $\pi_{PMF}$ are comoving variables which stay constant of time\cite{2002PhRvD..65l3004M,2000PhRvD..61d3001D,2007PhRvD..75b3002K,2004PhRvD..70d3011L}.

The vector mode evolution equation becomes,
\begin{eqnarray}
\ddot{H}_T^{(1)}+2\frac{\dot{a}}{a}\dot{H}_T^{(1)}&=&8\pi Ga^2(p_f\pi_f^{(1)}+p_{\gamma}\pi_{PMF}^{(1)})\nonumber\\
&\equiv&8\pi Ga^2\frac{\pi^{(1)}_{total}}{a^4}~~.
\label{eq108}
\end{eqnarray}
For relativistic components, the pressure $p_f$ is proportional to $a^{-4}$. Then, assuming a scaling relation, 
\begin{equation}
\pi_{total}^{(1)}\propto a^{\alpha}~~,
\label{eq120}
\end{equation}
and defining the shear as,
\begin{eqnarray}
\sigma^{(m)}=-\dot{H}_T^{(m)}/k,
\end{eqnarray}
 we get a simple solution to the vector mode equation  Eq.~(\ref{eq108}),
\begin{eqnarray}
\sigma^{(1)}=c_1a^{\alpha-1}+c_2a^{-2}~~ {\rm (Radiation-dominated)}\nonumber\\
\sigma^{(1)}=c_1a^{\alpha-1.5}+c_2a^{-2} ~~{\rm (Matter-dominated)}~.
\label{eq110}
\end{eqnarray}
Here, the power spectral index $\alpha$ and coefficients $c_1$ and $c_2$ depend on whether one is treating  the radiation or matter dominated epoch, as discussed later. 

The shear affects the CMB spectrum through Eqs. (\ref{eq103}) and (\ref{eq104}).
If there is no source of anisotropic stress, i.e. $\pi^{(1)}_f=\pi^{(1)}_{PMF}=0$, Eq.~(\ref{eq110}) has a simple decaying solution $\sigma^{(1)}=c_2a^{-2}$. This means that there are no vector CMB anisotropies if there is no anisotropic stress because the potential decays rapidly. This is the reason why the vector mode is usually ignored in CMB theory. However, if there is a source of anisotropic stress, $\sigma^{(1)}$ decreases slowly.

The tensor mode evolution equation is
\begin{eqnarray}
\ddot{H}_T^{(2)}+2\frac{\dot{a}}{a}\dot{H}_T^{(2)}+k^2H_T^{(2)}&=&8\pi Ga^2(p_f\pi_f^{(2)}+p_{\gamma}\pi_{PMF}^{(2)})\nonumber\\
&\equiv&8\pi Ga^2\frac{\pi^{(2)}_{total}}{a^4}~~.
\label{eq109}
\end{eqnarray}
Note, that unlike the vector mode evolution  [Eq.  (\ref{eq108})],  there is a linear term, $k^2H_T^{(2)}$, on the  l.h.s.~of Eq.~(\ref{eq109}).
 Assuming that all scales are outside of the horizon, we get a solution for the vector mode which is similar to Eq. (\ref{eq110}), i.e.
\begin{eqnarray}
\sigma^{(2)}=c_1a^{\alpha-1}+c_2a^{-2} ~~{\rm (Radiation-dominated)}\nonumber\\
\sigma^{(2)}=c_1a^{\alpha-1.5}+c_2a^{-2} ~~{\rm (Matter-dominated)}~.
\label{eq111}
\end{eqnarray}
This implies that, if there is a primordial magnetic field, $H_T^{(2)}$ grows even if there is no initial $H_T^{(2)}$. 
This passive mode was studied in detail  in Ref.~\cite{2004PhRvD..70d3011L}. However, we do not consider this mode further here, because we are interested only in the neutrino mass effects.

The essential effects  of massive neutrinos have been analyzed previously in Refs.~\cite{1996ApJ...467...10D,2005PhRvD..71d3001I,2006PhR...429..307L}. There are two primary effects.   First is the free-streaming effect. Outside the horizon, there is no free streaming effect because the $qk/ \epsilon$ term on the r.h.s. of Eq.~(\ref{eq101}) is negligible. Of course, the $\Theta$s do "free stream"  to neighboring hierarchical equations once a scale enters the horizon. This free-streaming effect , however, decreases as neutrinos become non-relativistic because 
 the $qk/ \epsilon$ term on the r.h.s. of Eq. (\ref{eq101}) becomes negligible again. 

The second effect is the change of anisotropic stress at the epoch when massive neutrinos become nonrelativistic.  Although anisotropic stress will decrease by the effect of neutrino mass, this change is very small without a PMF. On the other hand, the existence of the PMF affects drastically  the evolution of the anisotropic stress. We can understand this clearly from a comparison of the initial conditions for massless and massive neutrinos with and without a PMF. 

\subsection{Initial conditions for massless neutrinos}
We can derive initial conditions for massless neutrinos from the assumption that there are no radiative vorticity and anisotropic stress at very early times, i.e.~$\pi^{(m)}_{total}=0$. The result is \cite{2004PhRvD..70d3518L,2004PhRvD..70d3011L}
\begin{eqnarray}
\pi^{(1)}_{\nu}&\equiv& \sqrt[]{\mathstrut P(k)}\hat{\pi}^{(1)}_{\nu} \nonumber\\
&=&-\pi_{PMF}^{(1)}\frac{R_{\gamma}}{R_{\nu}}\Bigl(1-\frac{45}{14}\frac{(k\eta)^2}{4R_{\nu}+15}\Bigr)-\sigma_{i}^{(1)}\frac{2k\eta}{R_{\nu}}\nonumber\\
\pi^{(2)}_{\nu}&\equiv& \sqrt[]{\mathstrut P(k)}\hat{\pi}^{(2)}_{\nu} \nonumber\\
&=&-\pi_{PMF}^{(2)}\frac{R_{\gamma}}{R_{\nu}}\Bigl(1-\frac{15}{14}\frac{(k\eta)^2}{4R_{\nu}+15}\Bigr)+H_{T,i}^{(2)}\frac{4(k\eta)^2}{4R_{\nu}+15}~~,\nonumber\\
\label{eq112}
\end{eqnarray}
where $P(k)$ represents the power spectrum, $\hat{\pi}_{\nu}^{(m)}$ is the neutrino anisotropic stress normalized by $\sqrt[]{\mathstrut P(k)}$, $R_{\gamma}\equiv \rho_{\gamma}/\rho_{R}$, $R_{\nu}\equiv \rho_{\nu}/\rho_{R}$ and $\rho_{R}=\rho_{\gamma}+\rho_{\nu}$. Here $\sigma^{(1)}_{i}$ and $H_{T,i}^{(2)}$ are the initial conditions for the primary mode.  These  are taken to be initially turned off in the PMF mode. Hereafter, any quantities denoted with a {\it hat} (e.g.~$\hat{X}$) are the variables normalized by the square root of power spectrum, $\sqrt[]{\mathstrut P(k)}$. They are very useful variables for the illustration of  the effects of massive neutrinos to be given in Figs. \ref{fig:2} and \ref{fig:3}.

\subsection{Initial conditions for neutrinos with finite mass}
The initial conditions for massive neutrinos 
are a bit different from those of massless neutrinos. Expanding $q/\epsilon\simeq 1-m^2a^2/(2q^2)$,  Eq. (\ref{eq100}) can be expressed as 
\begin{eqnarray}
\pi_h^{(m)}&=& a^{-4}\frac{const.}{p_h}\int q^3dq(1-\frac{1}{2}\frac{m_{\nu}^2a^2}{q^2})\bar{f}\Theta_{h2}^{(m)}\nonumber\\
&\simeq&\pi_{\nu}^{(m)}(1-\frac{1}{2}\frac{5}{7\pi^2}H_0^2\Omega_{R}m_{\nu}^2\eta^2)~~.
\label{eq119}
\end{eqnarray}
We here define the effective wave number $k_{eff}$ by
\begin{eqnarray}
&&k_{eff}^{2}=k^2+k_m^2~~,\\
&&k_m=\sqrt{\mathstrut \frac{1}{2}\frac{5}{7\pi^2}H_0^2\Omega_{R}\frac{4R_{\nu}+15}{c}m_{\nu}^2}~~,
\label{eq121}
\end{eqnarray}
where $c=45/14$ for the vector mode ($m=1$), $c=15/14$ for the tensor mode ($m=2$) and $\Omega_R$ is a density parameter
of radiation when all neutrinos are relativistic.
Inserting Eq. (\ref{eq112}) into Eq. (\ref{eq119}), we obtain
the initial conditions for the total anisotropic stress $\pi_{total}^{(m)}$ as
\begin{eqnarray}
\pi_{total}^{(1)}&\equiv&\sqrt[]{\mathstrut P(k)}\hat{\pi}_{total}^{(1)}\nonumber\\
&\simeq&\pi_{PMF}^{(1)}\frac{45}{14}\frac{(k_{eff}\eta)^2}{4R_{\nu}+15}\frac{p_{\gamma}}{a^{-4}}-\sigma_{i}^{(1)}\frac{2k\eta}{R_{\nu}}\frac{p_{\nu}}{a^{-4}}\nonumber\\
\pi_{total}^{(2)}&\equiv&\sqrt[]{\mathstrut P(k)}\hat{\pi}_{total}^{(2)}\nonumber\\
&\simeq& \pi_{PMF}^{(2)}\frac{15}{14}\frac{(k_{eff}\eta)^2}{4R_{\nu}+15}\frac{p_{\gamma}}{a^{-4}}+H_{T,i}^{(2)}\frac{4(k\eta)^2}{4R_{\nu}+15}\frac{p_{\nu}}{a^{-4}}~~.\nonumber\\
\label{eq114}
\end{eqnarray}
This condition is also valid for massless neutrinos by setting $m_{\nu}=0$, for which $k_{eff}=k$.
Recall that the $\sigma_{i}^{(1)}$ and $H_{T,i}^{(2)}$ are the initial conditions for the primary mode, which should be turned off in the initial PMF mode. This equation clearly shows that the neutrino mass effects correlate with the PMF to  leading order in  $\eta^2$ and would affect strongly the PMF mode of the CMB temperature and polarization anisotropies. Note, that the expressions in  Eq.~(\ref{eq114}) are only valid for the epoch before the massive neutrinos become nonrelativistic, i.e.~$a < a_{nr}$, which is given by \cite{2005PhRvD..71d3001I}
\begin{eqnarray}
a_{nr}&=&\frac{T_{\nu}}{T_{\nu,nr}}\nonumber\\
&=&\frac{1.5\times 10^{-3}}{\sum m_{\nu}/{\rm eV}}~~,
\end{eqnarray}
Thus,  $a_{nr}=0.84\times 10^{-3}$ for neutrino mass of  $\sum m_{\nu}=1.8{\rm eV}$. This scale factor $a_{nr}$ is comparable with $a_{rec}=9.2\times 10^{-4}$ at recombination, but larger than $a_{eq}\approx 10^{-4}$.

\subsection{Evolution of anisotropic stress}
From Eq.~(\ref{eq114})
the total anisotropic stress, $\hat{\pi}_{total}^{(m)}$, on the r.h.s. of Eqs. (\ref{eq108}) and (\ref{eq109}) grows as $\hat{\pi}_{total}\propto \eta^2$ for both  massless and massive neutrinos as long as the perturbation is outside of the horizon and neutrinos are relativistic. In the massless neutrino case, $|\hat{\pi}_{\nu}^{(m)}|$ decreases as time goes on and damps with some oscillations once a perturbation of wave number $k$ enters the horizon. Then the PMF anisotropic stress $|\hat{\pi}_{PMF}^{(m)}|$ dominates $|\hat{\pi}_{total}^{(m)}|$ at later epochs because $\hat{\pi}_{PMF}^{(m)}$ remains  constant. On the other hand, in massive neutrino case, the neutrino anisotropic stress $|\hat{\pi}_{h}^{(m)}|$ decrease simultaneously at all scales of the perturbations if the wave number $k$ satisfies $k \ll k_{m}$, and $|\hat{\pi}_{PMF}^{(m)}|$ quickly dominates $|\hat{\pi}_{total}^{(m)}|$. For perturbations at smaller scales with $k\gg k_m$, however, the neutrino mass effect is negligible and perturbations grow as if neutrinos have no mass. These features of massive neutrinos make large differences in the CMB.

\subsection{PMF power spectrum}
Before closing this section we make a note on the power spectrum $P(k)$ used in Eqs. (\ref{eq112}) and (\ref{eq114}).
In the usual CMB theory, the power spectrum is
\begin{eqnarray}
k^3P(k)=A_sk^{n_s-1},
\end{eqnarray}
where $A_s$ is the scalar amplitude and $n_s$ is the scalar power spectral index.
In the present calculations we need to use the power spectrum for the PMF \cite{2004PhRvD..70d3518L,2004PhRvD..70d3011L,2002PhRvD..65l3004M,2007PhRvD..75b3002K,2008arXiv0801.2572Y,2006ApJ...646..719Y} which can be written approximately as 
\begin{eqnarray}
k^3P(k)\propto k^{2n_B+6}~~.
\label{eq116}
\end{eqnarray}
It is to be noted that  the accurate formula \cite{2008arXiv0801.2572Y,2006ApJ...646..719Y} of this rank of a single power spectrum has different forms for the scalar, vector and tensor modes for the PMF as a function of $k$ and $k_C$ which is 
the cutoff wave number. See Ref. \cite{2006ApJ...646..719Y} and \cite{2008arXiv0801.2572Y} for more details.

\section{Results and Discussion}
We have applied these equations which include a finite neutrino mass to the CMB anisotropy code  {\it CAMB} \cite{Lewis:1999bs}.
As an illustration of  the CMB anisotropies in the presence of a PMF we consider an example with a fixed amplitude and spectral index for the PMF power spectrum.  Ultimately,  one would hope to be able to deduce the amplitude and spectral index from fits to the polarization spectrum.
For our purposes, however, we assume that the PMF is generated before the nucleosynthesis epoch. In that case,
the CMB anisotropies at high multipole places an upper limit on the magnetic field of $B_{\lambda} < 4.7{\rm nG}$ \cite{2006ApJ...646..719Y}, and a nearly scale invariant spectrum is preferred \cite{2006ApJ...646..719Y,2002PhRvD..65b3517C}. Therefore,  in the present work, we  fixed the PMF at a field strength of  $B_{\lambda}=4.7{\rm nG}$, and adopt a spectral index of $n_B=-2.9$. As a representative neutrino mass,  we choose a value of  $\sum m_{\nu}=1.8$ eV which is the upper limit from the scalar mode analysis, $\sum m_{\nu}<1.8$ eV, deduced in \cite{2007ApJS..170..377S},  and is near the upper limit deduced from the WMAP-3yr analysis, $\sum m_{\nu}<2$ eV, \cite{2006PhRvD..74b7302F}. Other parameters are taken at the best fit values from the combined CMB, SNIa, LSS  analysis from the WMAP-3yr data \cite{2007ApJS..170..377S}, except that the best fit value for $\Omega_{CDM}$ is replaced with  $\Omega_{CDM}-\Omega_{\nu}(1.8{\rm eV})$ in order to maintain  a flat universe model.

\subsection{Power spectra}
\begin{figure*}
\includegraphics[width=5cm,angle=270]{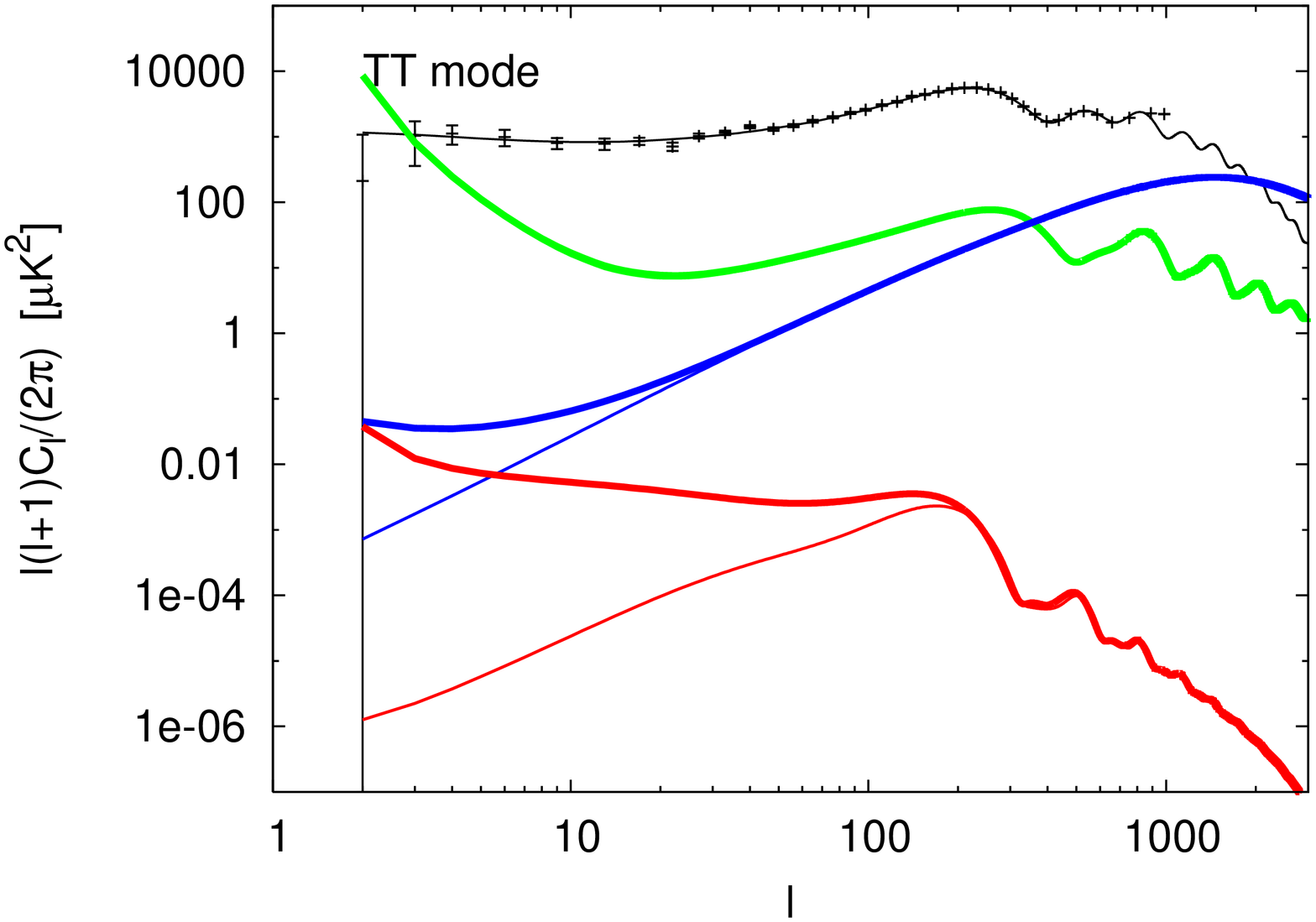}
\includegraphics[width=5cm,angle=270]{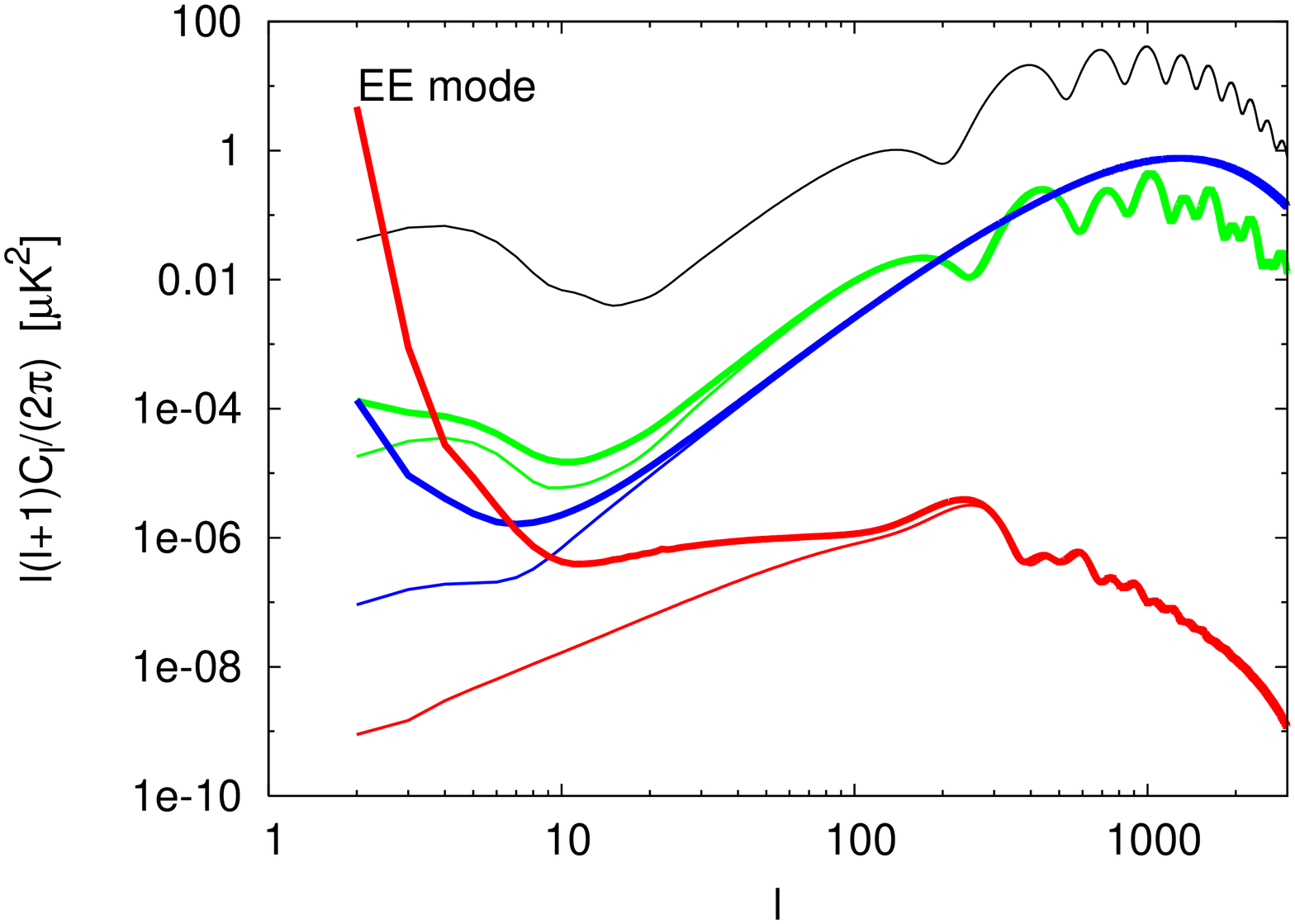}
\includegraphics[width=5cm,angle=270]{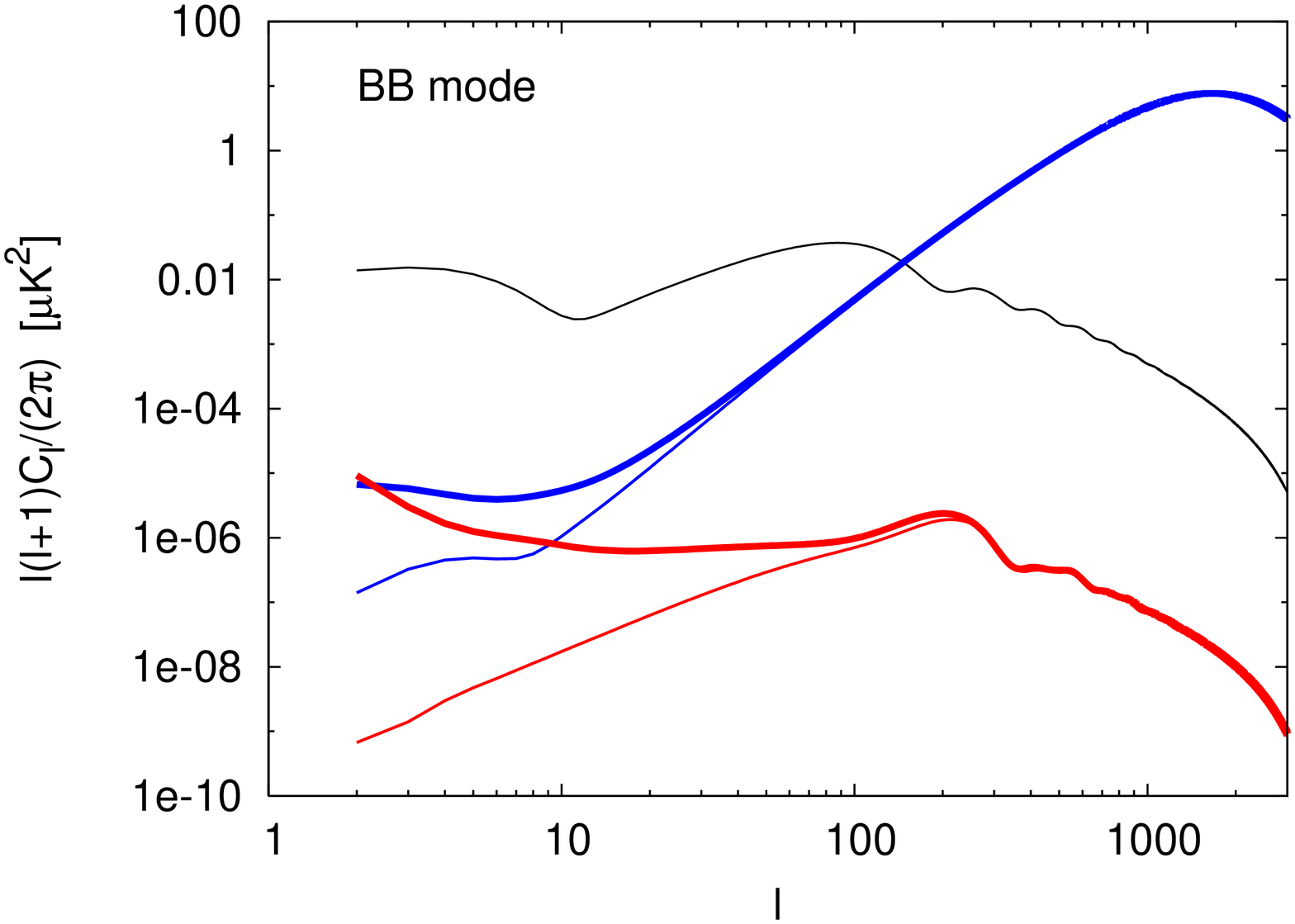}
\includegraphics[width=5cm,angle=270]{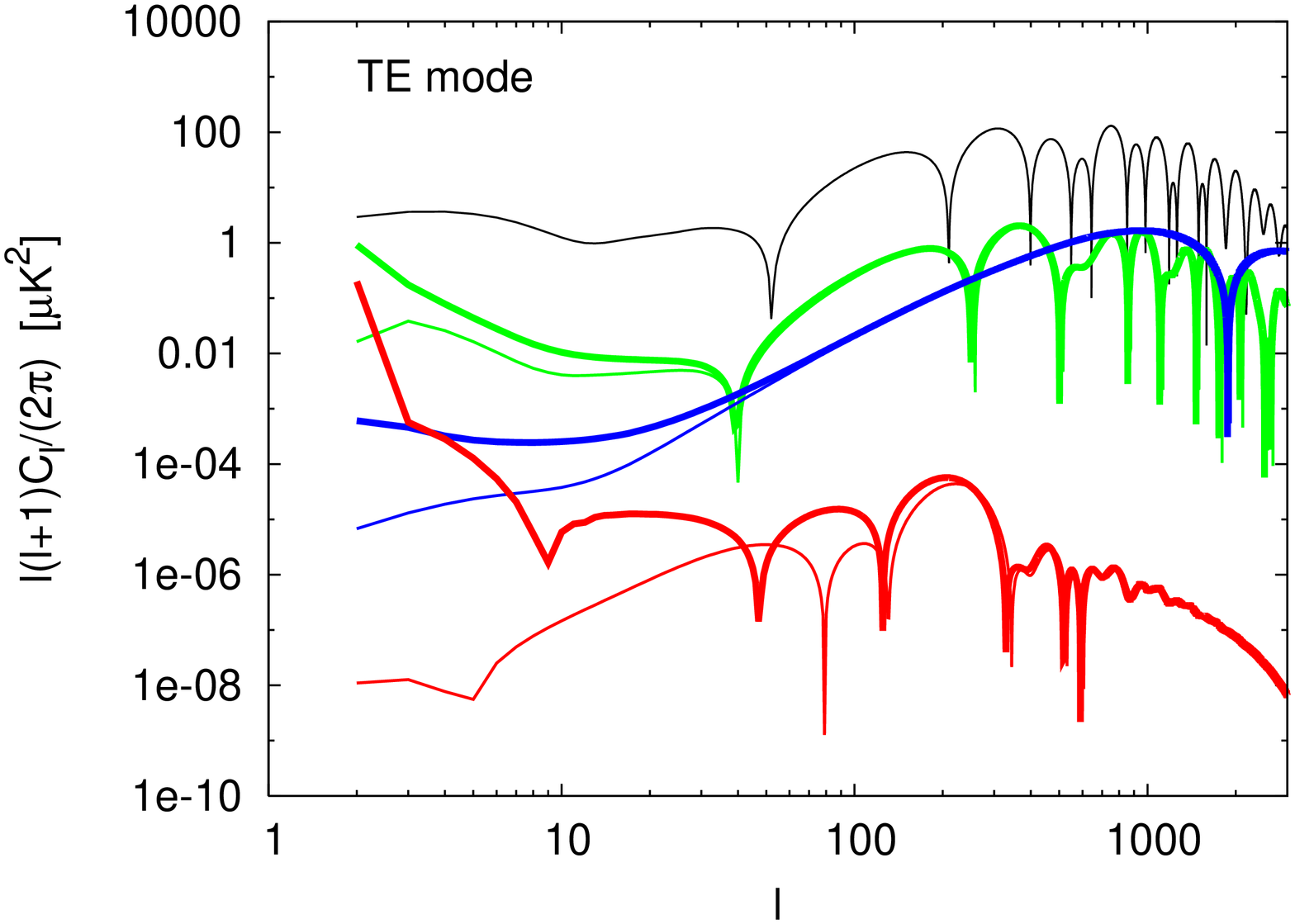}
\caption{The power spectrum of the TT, EE, BB, and TE  modes when a PMF is included. The black thin lines show the primary spectra, the scalar mode for the TT, EE, and TE mode, and the tensor mode for the BB mode.  For this figure the amplitude ratio of the tensor-to-scalar mode has been set to 0.55. The other thin lines represent massless neutrino models.  The thick lines show a model with massive neutrinos for which, $\sum m_{\nu}=1.8{\rm~eV}$. Green lines are for the scalar mode, blue lines are for the vector mode, and red lines are for the tensor mode. The amplitude of 
the PMF is taken to be $B_{\lambda}=4.7{\rm~nG}$, and the spectral index is taken to be $n_B=-2.9$. Data points for the TT mode are from the  WMAP-3yr data.}
\label{fig:1}
\end{figure*}

 Fig.~\ref{fig:1} shows the power spectra for the scalar, vector and tensor, TT, EE, and TE  modes along with the vector and tensor power spectra for the BB mode  when a PMF is included. The upper most black lines, except for the regions $l^<_{\sim} 3$ for TT mode and $l^>_{\sim} 100$ for BB mode, show the primary spectra, and the other thin and thick lines represent models with massless and massive neutrinos, respectively, which include a PMF. 

Massive neutrinos have little effect on the scalar TT power spectra as evidenced by the fact that the  thin green line and thick green line are almost indistinguishable from each other. The only effect of massive neutrinos in the scalar TT mode is a 3\% enhancement for  the quadrapole ($l=2$) fluctuation. Also, the massive neutrinos make almost no difference in the vector (blue) and tensor (magenta) modes except at  lower multipoles $l^<_\sim 100$.
There, an  excess of power at low $ l$  is caused  by an increase in the  shear as we now discuss.

\subsection{Analysis of shear}
To aid in this discussion, Figs.  \ref{fig:2} and \ref{fig:3} show the absolute values of the shear $\hat{\sigma}^{(m)}$, total anisotropic stress $\hat{\pi}^{(m)}$, anisotropic scattering source $\hat{P}^{(m)}$, and the tensor metric perturbation $\hat{H}_T^{(2)}$.  These quantities  are normalized by the square root of power spectrum, $\sqrt[]{\mathstrut P(k)}$, as was defined below Eq. (\ref{eq112}). They are plotted as a function of the scale factor $a$ for the vector (Fig. \ref{fig:2}) and tensor (Fig. \ref{fig:3}) modes. These quantities are shown for five different scales of the wave number $k=10^{-2},~5\times 10^{-3},~10^{-3},~10^{-4},~10^{-5}~{\rm Mpc^{-1}}$ from top to bottom in each panel of these figures.

 In each figure the l.h.s. panels display the models for massless neutrinos, and the r.h.s. panels display the models for massive neutrinos. The two upper-most lines for the larger wave numbers (smaller scales)  ($k=10^{-2}$ and $5\times 10^{-3}~{\rm Mpc^{-1}}$) are very similar to each other for both  massless and massive neutrinos. However, the other larger scales $k\le 10^{-3}{\rm Mpc^{-1}}$ show a   totally different evolution for massless vs. massive neutrinos. We understand the reason for similarity and difference of these quantities between the two models of massless and massive neutrinos as we now describe.

\subsection{Growth of perturbations}

\begin{figure*}
\includegraphics[width=12cm,angle=270]{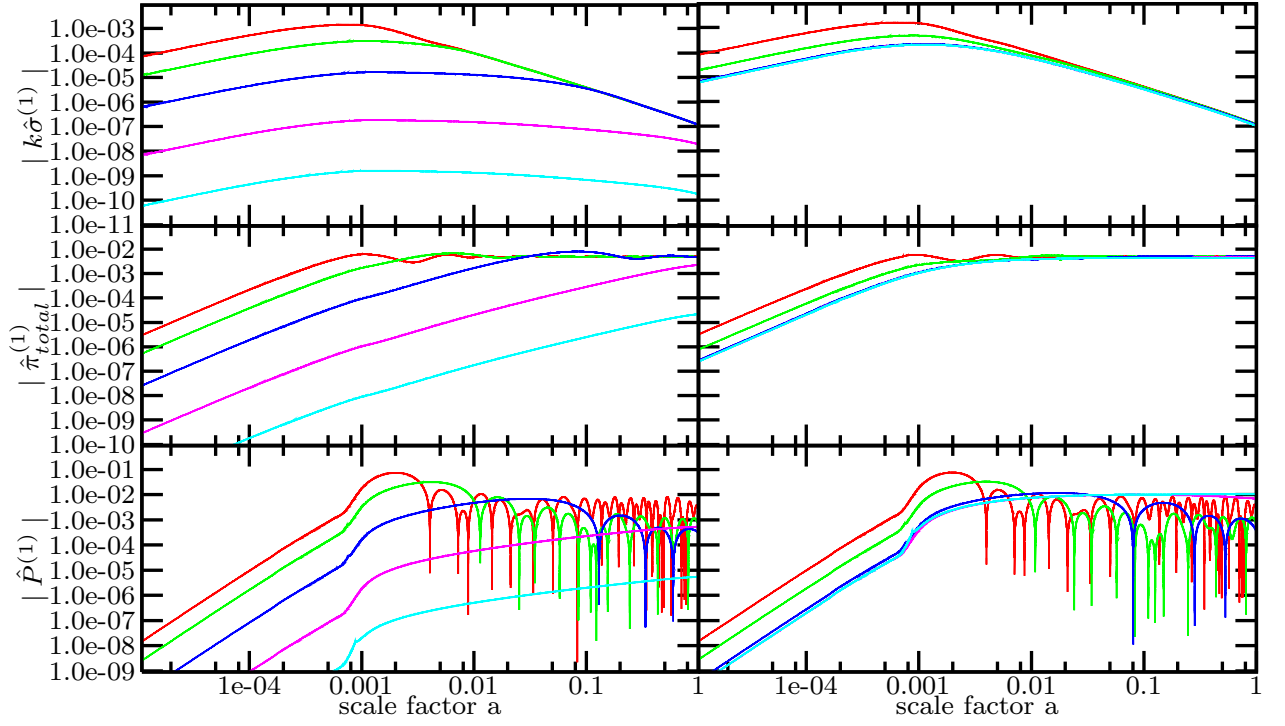}
\caption{Absolute values of the shear $k\hat{\sigma}^{(1)}$, total anisotropic stress $\hat{\pi}_{total}^{(1)}$, and the anisotropic scattering source $\hat{P}^{(1)}$, normalized by the square root of power spectrum, for the  vector mode as a function of the scale factor $a$. The figures on the left hand side are for $\sum m_{\nu}=0{\rm~eV}$, and the right hand side figures are  for $\sum m_{\nu}=1.8{\rm~eV}$. Each line corresponds to a  different scale, i.e.
 red: $k=0.01$, green: $k=0.005$, blue: $k=0.001$, magenta: $k=0.0001$, light blue: $k=0.00001$ ($\rm Mpc^{-1}$). 
 }
\label{fig:2}
\end{figure*}

\begin{figure*}
\includegraphics[width=12cm,angle=270]{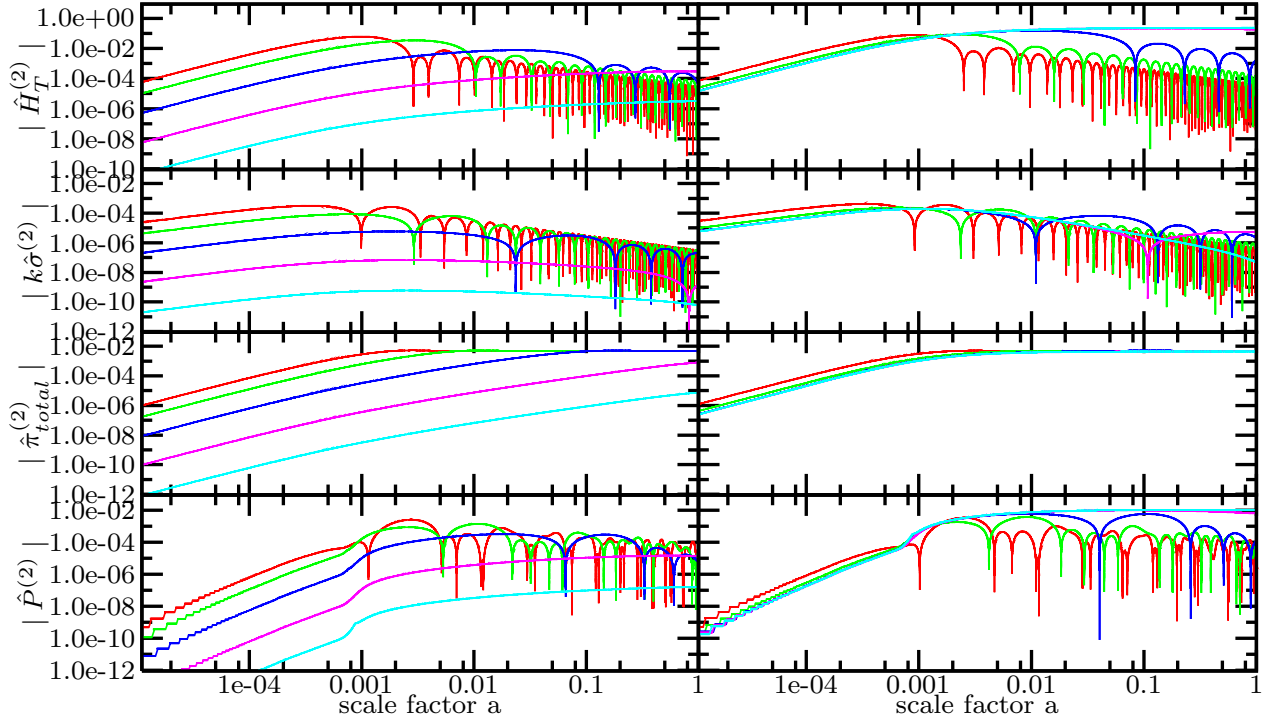}
\caption{Absolute values of the tensor-mode metric perturbation $\hat{H}_T^{(2)}$, the shear $k\hat{\sigma}^{(2)}$, the total anisotropic stress $\hat{\pi}_{total}^{(2)}$, and the anisotropic scattering source $\hat{P}^{(2)}$, normalized by the square root of power spectrum,  as a function of the scale factor $a$. The figures on the left hand side are for  $\sum m_{\nu}=0 {\rm~eV}$, and the right hand side figures are  for $\sum m_{\nu}=1.8{\rm~eV}$. Each line corresponds to a  different scale, i.e.
 red: $k=0.01$, green: $k=0.005$, blue: $k=0.001$, magenta: $k=0.0001$, light blue: $k=0.00001$ ($\rm Mpc^{-1}$).
 }
\label{fig:3}
\end{figure*}

In the case when a perturbation is outside of the horizon, $k \eta\ll 1$,
we know from Eq. (\ref{eq114}) that the normalized total anisotropic stress $\hat{\pi}_{total}^{(m)}$ grows quadratically ($\propto \eta^2$). Using the fact that $\eta\propto a$ during the  radiation dominated epoch ($a<a_{eq}$) and that $\eta \propto a^{1/2}$ during the  matter dominated epoch ($a>a_{eq}$), it is clear that $\hat{\pi}_{total}^{(m)}\propto a^2$ (i.e. $\alpha=2$) in the  radiation dominated epoch, and $\hat{\pi}_{total}^{(m)}\propto a$ (i.e. $\alpha=1$) during the  matter dominated epoch.  This behavior is apparent  in 
Figs.  \ref{fig:2} and \ref{fig:3}. 

In the same  way we can understand the growth of shear. From Eqs. (\ref{eq110}) and (\ref{eq111}) and $\alpha$ obtained from the above, the growth of the shear becomes $\hat{\sigma}^{(m)}\propto a$ during the radiation dominated epoch and $\hat{\sigma}^{(m)}\propto a^{-0.5}$ in the  matter dominated epoch, as is also evident in  Figs.  \ref{fig:2} and \ref{fig:3}.

As time goes on, the anisotropic stress of the neutrinos decreases, and the PMF anisotropic stress becomes dominant, as can be seen in Eq. (\ref{eq114}). 
However, once a perturbation enters the horizon, hierarchical mixing sets in, and the initial condition Eq. (\ref{eq114}) is no longer valid. It is not easy to derive an analytic solution for the anisotropic stress. However, we  know from numerical calculations that neutrino anisotropic stress undergoes damped oscillations, and $\hat{\pi}_{total}^{(m)}$ remains constant asymptotically as $\propto a^0$ as shown in Figs.  \ref{fig:2} and \ref{fig:3}. Having this result, and applying  $\alpha\approx 0$ to Eqs. (\ref{eq110}) and (\ref{eq111}), we can expect $\hat{\sigma}^{(m)}=c_1a^{\alpha-1.5}+c_2a^{-2}\approx c_1a^{\alpha-1.5}\propto a^{-1.5}$ during the  matter dominated era. This damping power spectral index $-1.5$ is in good agreement with the slope of $k\hat{\sigma}^{(m)}$ seen in Figs.  \ref{fig:2} and \ref{fig:3}.

If the neutrinos have  mass, the evolution of the shear $k\hat{\sigma}^{(m)}$, the total anisotropic stress $\hat{\pi}_{total}^{(m)}$, and the anisotropic scattering source $\hat{P}^{(m)}$ drastically change as displayed on the r.h.s.~panels of Figs.  \ref{fig:2} and \ref{fig:3}.  In this case of finite mass neutrinos, initial conditions of the form of Eq.~(\ref{eq114}) are valid when perturbations are outside the horizon and when neutrinos behave relativistically, i.e.~when $a < a_{nr} \approx 0.84 \times 10^{-3}$.

There is a distinctive feature of nearly equivalent evolution for  these quantities especially at larger scales, $k_\sim^< 10^{-3}{\rm Mpc^{-1}}$. When one takes $\sum m_{\nu}=1.8 {\rm eV}$, the critical wave number $k_m$ of Eq. (\ref{eq121}) is $k_m=3.5\times 10^{-3}{\rm Mpc^{-1}}$ for the vector mode and $k_m=6.0\times 10^{-3}{\rm Mpc^{-1}}$ for the tensor mode. Therefore, $k_{eff}\approx k$ for smaller scales, $k>k_m$, and $k_{eff}\approx k_m=const.$ for larger scales, $k<k_m$. 
Putting these conditions into Eq. (\ref{eq114}),  a similar evolution to that of massless neutrino models is expected for smaller scales $k>k_m$.  However, the  evolution becomes almost degenerate for larger scales $k<k_m\sim 10^{-3}{\rm Mpc^{-1}}$.
This is the reason for the drastic change from massless neutrinos to massive neutrinos in $k\hat{\sigma}^{(m)}$, $\hat{\pi}_{total}^{(m)}$ and $\hat{P}^{(m)}$ for larger scales with smaller wave number $k<k_m$. 

The effect of such changes brought on by   massive neutrinos  is even more dramatic at low multipoles in the CMB power spectrum. The multipole $l_m$ corresponding to the scale $k_m$ where the evolution becomes almost degenerate is 
\begin{eqnarray}
l_m\sim k_m\eta_0~~,
\end{eqnarray}
where $\eta_0\sim 14{\rm Gpc}$ in the standard $\Lambda$CDM model. Since $k_m$ is known to be $3.5\times 10^{-3}{\rm Mpc^{-1}}$ and $6.0\times 10^{-3}{\rm Mpc^{-1}}$ for the vector and tensor modes, respectively, the critical multipole $l_m$ turns out to be 
\begin{eqnarray}
&&l_m\sim 50~~{\rm (Vector~mode)}~,\nonumber \\
&&l_m\sim 85~~{\rm (Tensor~mode)}~.
\end{eqnarray} 
Hence, the CMB power spectrum for lower mutipoles $l<l_m$, corresponding to smaller $k<k_m$, is expected to stay at the same value, which is in reasonable agreement with the calculated results shown in Fig. \ref{fig:1}.
 In fact, there are also neutrino effects in the scalar mode at low $l $. However, they are ambiguous because of confusion from the integrated Sachs-Wolfe effect which is large when a  PMF is present.

\subsection{EE and BB Modes}
\begin{figure*}
\includegraphics[width=5cm,angle=270]{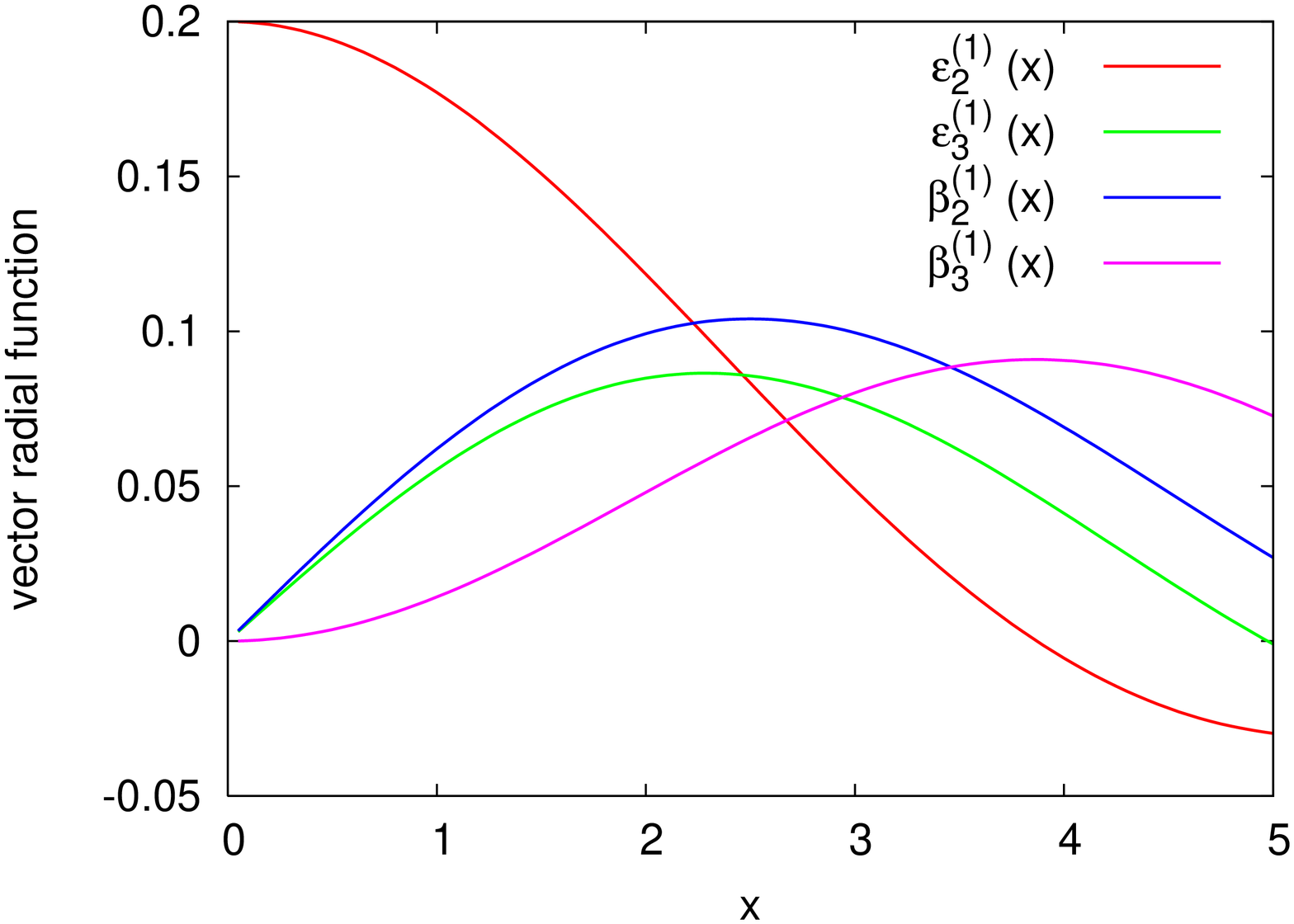}
\includegraphics[width=5cm,angle=270]{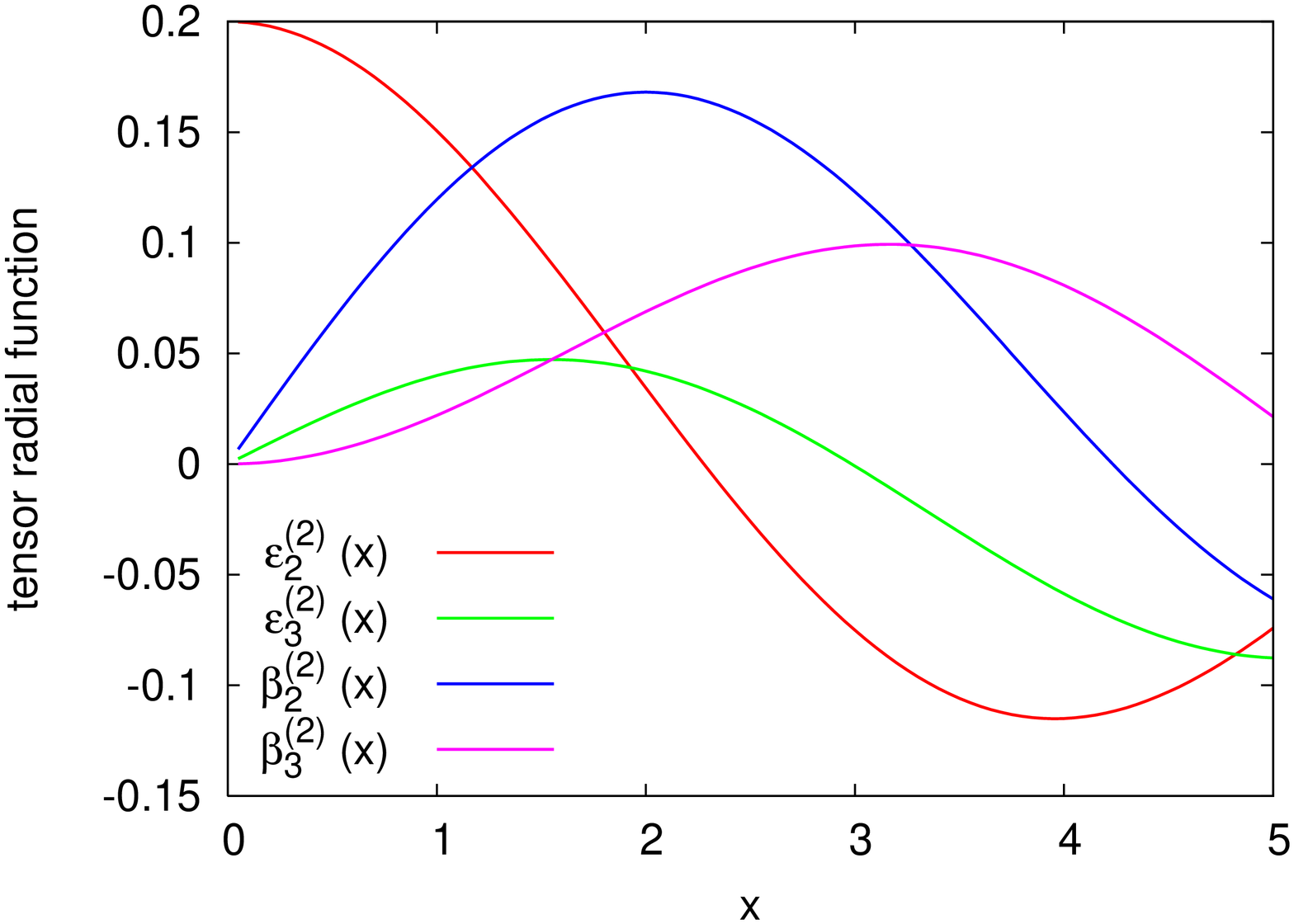}
\caption{Radial E function $\epsilon_l^{(m)}(x)$ and radial B function $\beta_l^{(m)}(x)$, where $l=2,3$ and $m=1,2$. The left hand side is for the vector mode, and the right hand side is for the  tensor mode.}
\label{fig:4}
\end{figure*}

The neutrino mass effect on the EE mode is larger than that of the BB mode. This difference is caused by the nature of the radial E and B functions $\epsilon_l^{(m)}(x)$ and $\beta_l^{(m)}(x)$. The behavior of these functions at smaller $x\le 5$ is plotted in Fig.~\ref{fig:4} for $l=2,3$ and $m=1,2$. From the definition of the radial E and B functions in Ref. \cite{1997PhRvD..56..596H}, their leading order term behaves as $\epsilon_l^{(m)}(x) \propto x^{l-2}$ and $\beta_l^{(m)}(x)\propto x^{l-1}$ at $x\sim 0$, where we have used a form of spherical bessel function $j_l(x)\propto x^l$ at $x\sim 0$. Only the $\epsilon_2^{(m)}(x)$ function is finite at $x=0$, i.e. $\epsilon_2^{(m)}\rightarrow 0.2$ as $x\rightarrow 0$, while the other radial functions all vanish, $\rightarrow 0$ at $x\rightarrow 0$. This finiteness of $\epsilon_2^{(m)}(0)$, combined with a nearly scale invariant power spectrum for the PMF, i.e. $k^3P(k)\propto k^{2n_B+6}\sim k^0$ for $n_B\sim -3$ from  simple approximation Eq. (\ref{eq116}), causes excess power for both the vector ($m=1$) and tensor ($m=2$) components in the EE and TE modes for $l=2$, as we discuss below. 

Using the fact that the visibility function is approximated by a delta function $\dot{\tau}e^{-\tau}\simeq \delta(\eta-\eta_{rec})$ and that $\hat{P}^{(m)}\propto k_{eff}^2$ for scales outside the horizon,
we can estimate the CMB power spectrum of Eq. (\ref{eq115}) as
\begin{eqnarray}
&&(2l+1)^2C_l^{EE(m)}\nonumber\\
&&~~~~~~\propto \int \frac{dk}{k}k^{2n_B+6}k_{eff}^{4}\epsilon_l^{(m)2}(k(\eta_0-\eta_{rec}))~~.
\label{eq117}
\end{eqnarray}
If neutrinos are light enough to be relativistic at recombination, Eq. (\ref{eq117}) is a very good approximation for scales outside of the horizon for $l_\sim^<100$. Note that this approximation is also valid in our calculation because $a_{nr}\sim a_{rec}$.  

For the massless neutrino case, $m_{\nu}=0$, we can insert $k_{eff}=k$ into Eq. (\ref{eq117}). We then obtain
\begin{eqnarray}
(2l+1)^2C_2^{EE(m)}\propto \int dk k^{2n_B+9}\epsilon_2^{(m)2}(k(\eta_0-\eta_{rec}))~~.
\label{eq118}
\end{eqnarray}
This converges to a finite value for all multipoles $l$ even when one takes a nearly scale invariant power spectrum $P(k)$ with $n_B\sim -3$. Our calculated CMB power spectra for the  EE mode displayed in Fig. \ref{fig:1} does  not show any excess at $l\le 10$ in the massless case. However, if the neutrinos have  mass, $k_{eff}=\sqrt{\mathstrut k^2+k_m^2}\simeq k_m$ on very large scales, and Eq. (\ref{eq117}) becomes 
\begin{eqnarray}
&&(2l+1)^2C_l^{EE(m)}(m_{\nu}\neq 0)\nonumber\\
&&~~~~~\propto \int dkk^{2n_B+5}k_{m}^{4}\epsilon_l^{(m)2}(k(\eta_0-\eta_{rec}))\nonumber\\
&&~~~~~\propto\int dk k^{2(n_B+3)+2l-5}~~,
\label{eq122}
\end{eqnarray}
where we have set  $\epsilon_l^{(m)}(x)\propto x^{l-2}~~(x\sim 0)$ in the integrand on the r.h.s. When one takes a nearly  scale invariant power spectrum ($n_B\sim -3$ ) for the PMF [$P(k)$ of Eq. (\ref{eq116})], this integral has a logarithmic  infrared divergence  for the quadrapole $l=2$ term, although it is regular for the higher multipoles $l\ge 3$. In the present calculation, we have  set  $n_B=-2.9$, because  
it is impossible to set $n_B=-3$ \cite{2002PhRvD..65l3004M,2007PhRvD..75b3002K} although this nearly scale invariant spectrum is preferred \cite{2006ApJ...646..719Y,2002PhRvD..65b3517C}. This is the reason why our calculated EE mode shows  a huge (but finite) excess for $l=2$.

For these mechanisms the anisotropies of the tensor EE mode becomes 100 times larger than the primary power  spectrum. 
The ratio of massive to massless neutrinos, $C_{l,h}/C_{l,\nu}$, is a function of neutrino mass 
and spectral index $n_B$ as we can see from Eq.~(\ref{eq114}) and (\ref{eq122}). The ratio does not depend on 
the amplitude of magnetic field, $B_{\lambda}$, because $C_l$ varies as $C_l \propto B_{\lambda}^4$ \cite{2004PhRvD..70d3011L} in the same manner independently of the neutrino mass.
Keeping this in mind, since strong enhancement of the anisotropies of tensor EE mode 
 depends upon $B_{\lambda}$, $n_B$, and $\sum m_{\nu}$, it may be possible to observe this effect in the future and thereby place a strong constraint on all three of these quantities and on the neutrino mass in particular. 

There is however a large effect of cosmic variance in lower $l$, which
is proportional to $\sqrt[]{\mathstrut 2/(2l+1)}$ because of the finiteness of sampling \cite{1997PhRvD..55.7368K}.
This variance makes it difficult to clearly observe neutrino mass effect.
Even if we cannot obtain lower limit of power spectrum due to the cosmic variance,
we can still obtain the upper limit at lower $l$, from which we can constrain the upper limit of the neutrino mass.
There is also a high probability of observing the neutrino mass
effect in the BB mode if the gravitational wave is sufficiently weak, though we should
consider carefully the passive mode studied by Lewis~\cite{2004PhRvD..70d3011L} in BB mode.

\section{Conclusion}
In this work, we have expanded on earlier studies of the scalar CMB anisotropies in the presence of a PMF.  In particular we have derived new  vector and tensor mode equations in the presence of  massive neutrinos and a PMF.  We find a large  effect from a finite neutrino mass on the vector and tensor modes  when a PMF exists.  In particular, the effect of  massive neutrinos on the EE mode become comparable to the observed primary anisotropy. Therefore, if and when the polarization power spectrum is ever measured at low multipoles, the possibility may exist  to place a much stronger constraint on the sum of the neutrino masses than presently exists, though the effect of  
cosmic variance should be carefully taken into consideration.
\section{Acknowledgment}
D.G.Y. and K. I. acknowledge the support by Grants-in-Aid for JSPS Fellows.
This work has been supported in part by Grants-in-Aid for Scientific
Research (17540275) of the Ministry of Education, Culture, Sports,
Science and Technology of Japan, and the Mitsubishi Foundation.  This
work is also supported by the JSPS Core-to-Core Program, International
Research Network for Exotic Femto Systems (EFES).  Work at UND supported in part by the US Department of Energy under research grant DE-FG02-95-ER40934.

\newpage 
\bibliography{neutrino1-12}

\begin{thebibliography}{26}
\expandafter\ifx\csname natexlab\endcsname\relax\def\natexlab#1{#1}\fi
\expandafter\ifx\csname bibnamefont\endcsname\relax
  \def\bibnamefont#1{#1}\fi
\expandafter\ifx\csname bibfnamefont\endcsname\relax
  \def\bibfnamefont#1{#1}\fi
\expandafter\ifx\csname citenamefont\endcsname\relax
  \def\citenamefont#1{#1}\fi
\expandafter\ifx\csname url\endcsname\relax
  \def\url#1{\texttt{#1}}\fi
\expandafter\ifx\csname urlprefix\endcsname\relax\def\urlprefix{URL }\fi
\providecommand{\bibinfo}[2]{#2}
\providecommand{\eprint}[2][]{\url{#2}}

\bibitem[{\citenamefont{{Yamazaki} et~al.}(2006)\citenamefont{{Yamazaki},
  {Ichiki}, {Kajino}, and {Mathews}}}]{2006ApJ...646..719Y}
\bibinfo{author}{\bibfnamefont{D.~G.} \bibnamefont{{Yamazaki}}},
  \bibinfo{author}{\bibfnamefont{K.}~\bibnamefont{{Ichiki}}},
  \bibinfo{author}{\bibfnamefont{T.}~\bibnamefont{{Kajino}}}, \bibnamefont{and}
  \bibinfo{author}{\bibfnamefont{G.~J.} \bibnamefont{{Mathews}}},
  \bibinfo{journal}{\apj} \textbf{\bibinfo{volume}{646}}, \bibinfo{pages}{719}
  (\bibinfo{year}{2006}), \eprint{arXiv:astro-ph/0602224}.

\bibitem[{\citenamefont{{Lewis}}(2004{\natexlab{a}})}]{2004PhRvD..70d3518L}
\bibinfo{author}{\bibfnamefont{A.}~\bibnamefont{{Lewis}}},
  \bibinfo{journal}{\prd} \textbf{\bibinfo{volume}{70}},
  \bibinfo{pages}{043518} (\bibinfo{year}{2004}{\natexlab{a}}),
  \eprint{arXiv:astro-ph/0403583}.

\bibitem[{\citenamefont{{Lewis}}(2004{\natexlab{b}})}]{2004PhRvD..70d3011L}
\bibinfo{author}{\bibfnamefont{A.}~\bibnamefont{{Lewis}}},
  \bibinfo{journal}{\prd} \textbf{\bibinfo{volume}{70}},
  \bibinfo{pages}{043011} (\bibinfo{year}{2004}{\natexlab{b}}),
  \eprint{arXiv:astro-ph/0406096}.

\bibitem[{\citenamefont{{Mack} et~al.}(2002)\citenamefont{{Mack},
  {Kahniashvili}, and {Kosowsky}}}]{2002PhRvD..65l3004M}
\bibinfo{author}{\bibfnamefont{A.}~\bibnamefont{{Mack}}},
  \bibinfo{author}{\bibfnamefont{T.}~\bibnamefont{{Kahniashvili}}},
  \bibnamefont{and}
  \bibinfo{author}{\bibfnamefont{A.}~\bibnamefont{{Kosowsky}}},
  \bibinfo{journal}{\prd} \textbf{\bibinfo{volume}{65}},
  \bibinfo{pages}{123004} (\bibinfo{year}{2002}),
  \eprint{arXiv:astro-ph/0105504}.

\bibitem[{\citenamefont{{Kahniashvili} and
  {Ratra}}(2007)}]{2007PhRvD..75b3002K}
\bibinfo{author}{\bibfnamefont{T.}~\bibnamefont{{Kahniashvili}}}
  \bibnamefont{and} \bibinfo{author}{\bibfnamefont{B.}~\bibnamefont{{Ratra}}},
  \bibinfo{journal}{\prd} \textbf{\bibinfo{volume}{75}},
  \bibinfo{pages}{023002} (\bibinfo{year}{2007}),
  \eprint{arXiv:astro-ph/0611247}.

\bibitem[{\citenamefont{{Kuo} et~al.}(2004)\citenamefont{{Kuo}, {Ade}, {Bock},
  {Cantalupo}, {Daub}, {Goldstein}, {Holzapfel}, {Lange}, {Lueker}, {Newcomb}
  et~al.}}]{2004ApJ...600...32K}
\bibinfo{author}{\bibfnamefont{C.~L.} \bibnamefont{{Kuo}}},
  \bibinfo{author}{\bibfnamefont{P.~A.~R.} \bibnamefont{{Ade}}},
  \bibinfo{author}{\bibfnamefont{J.~J.} \bibnamefont{{Bock}}},
  \bibinfo{author}{\bibfnamefont{C.}~\bibnamefont{{Cantalupo}}},
  \bibinfo{author}{\bibfnamefont{M.~D.} \bibnamefont{{Daub}}},
  \bibinfo{author}{\bibfnamefont{J.}~\bibnamefont{{Goldstein}}},
  \bibinfo{author}{\bibfnamefont{W.~L.} \bibnamefont{{Holzapfel}}},
  \bibinfo{author}{\bibfnamefont{A.~E.} \bibnamefont{{Lange}}},
  \bibinfo{author}{\bibfnamefont{M.}~\bibnamefont{{Lueker}}},
  \bibinfo{author}{\bibfnamefont{M.}~\bibnamefont{{Newcomb}}},
  \bibnamefont{et~al.}, \bibinfo{journal}{\apj} \textbf{\bibinfo{volume}{600}},
  \bibinfo{pages}{32} (\bibinfo{year}{2004}), \eprint{arXiv:astro-ph/0212289}.

\bibitem[{\citenamefont{{Reichardt} et~al.}(2008)\citenamefont{{Reichardt},
  {Ade}, {Bock}, {Bond}, {Brevik}, {Contaldi}, {Daub}, {Dempsey}, {Goldstein},
  {Holzapfel} et~al.}}]{2008arXiv0801.1491R}
\bibinfo{author}{\bibfnamefont{C.~L.} \bibnamefont{{Reichardt}}},
  \bibinfo{author}{\bibfnamefont{P.~A.~R.} \bibnamefont{{Ade}}},
  \bibinfo{author}{\bibfnamefont{J.~J.} \bibnamefont{{Bock}}},
  \bibinfo{author}{\bibfnamefont{J.~R.} \bibnamefont{{Bond}}},
  \bibinfo{author}{\bibfnamefont{J.~A.} \bibnamefont{{Brevik}}},
  \bibinfo{author}{\bibfnamefont{C.~R.} \bibnamefont{{Contaldi}}},
  \bibinfo{author}{\bibfnamefont{M.~D.} \bibnamefont{{Daub}}},
  \bibinfo{author}{\bibfnamefont{J.~T.} \bibnamefont{{Dempsey}}},
  \bibinfo{author}{\bibfnamefont{J.~H.} \bibnamefont{{Goldstein}}},
  \bibinfo{author}{\bibfnamefont{W.~L.} \bibnamefont{{Holzapfel}}},
  \bibnamefont{et~al.}, \bibinfo{journal}{ArXiv e-prints}
  \textbf{\bibinfo{volume}{801}} (\bibinfo{year}{2008}), \eprint{0801.1491}.

\bibitem[{\citenamefont{{Mason} et~al.}(2003)\citenamefont{{Mason}, {Pearson},
  {Readhead}, {Shepherd}, {Sievers}, {Udomprasert}, {Cartwright}, {Farmer},
  {Padin}, {Myers} et~al.}}]{2003ApJ...591..540M}
\bibinfo{author}{\bibfnamefont{B.~S.} \bibnamefont{{Mason}}},
  \bibinfo{author}{\bibfnamefont{T.~J.} \bibnamefont{{Pearson}}},
  \bibinfo{author}{\bibfnamefont{A.~C.~S.} \bibnamefont{{Readhead}}},
  \bibinfo{author}{\bibfnamefont{M.~C.} \bibnamefont{{Shepherd}}},
  \bibinfo{author}{\bibfnamefont{J.}~\bibnamefont{{Sievers}}},
  \bibinfo{author}{\bibfnamefont{P.~S.} \bibnamefont{{Udomprasert}}},
  \bibinfo{author}{\bibfnamefont{J.~K.} \bibnamefont{{Cartwright}}},
  \bibinfo{author}{\bibfnamefont{A.~J.} \bibnamefont{{Farmer}}},
  \bibinfo{author}{\bibfnamefont{S.}~\bibnamefont{{Padin}}},
  \bibinfo{author}{\bibfnamefont{S.~T.} \bibnamefont{{Myers}}},
  \bibnamefont{et~al.}, \bibinfo{journal}{\apj} \textbf{\bibinfo{volume}{591}},
  \bibinfo{pages}{540} (\bibinfo{year}{2003}), \eprint{arXiv:astro-ph/0205384}.

\bibitem[{\citenamefont{{Xu} et~al.}(2006)\citenamefont{{Xu}, {Kronberg},
  {Habib}, and {Dufton}}}]{2006ApJ...637...19X}
\bibinfo{author}{\bibfnamefont{Y.}~\bibnamefont{{Xu}}},
  \bibinfo{author}{\bibfnamefont{P.~P.} \bibnamefont{{Kronberg}}},
  \bibinfo{author}{\bibfnamefont{S.}~\bibnamefont{{Habib}}}, \bibnamefont{and}
  \bibinfo{author}{\bibfnamefont{Q.~W.} \bibnamefont{{Dufton}}},
  \bibinfo{journal}{\apj} \textbf{\bibinfo{volume}{637}}, \bibinfo{pages}{19}
  (\bibinfo{year}{2006}), \eprint{arXiv:astro-ph/0509826}.

\bibitem[{\citenamefont{{Clarke} et~al.}(2001)\citenamefont{{Clarke},
  {Kronberg}, and {B{\"o}hringer}}}]{2001ApJ...547L.111C}
\bibinfo{author}{\bibfnamefont{T.~E.} \bibnamefont{{Clarke}}},
  \bibinfo{author}{\bibfnamefont{P.~P.} \bibnamefont{{Kronberg}}},
  \bibnamefont{and}
  \bibinfo{author}{\bibfnamefont{H.}~\bibnamefont{{B{\"o}hringer}}},
  \bibinfo{journal}{\apj} \textbf{\bibinfo{volume}{547}}, \bibinfo{pages}{L111}
  (\bibinfo{year}{2001}), \eprint{arXiv:astro-ph/0011281}.

\bibitem[{\citenamefont{{Wolfe} et~al.}(1992)\citenamefont{{Wolfe}, {Lanzetta},
  and {Oren}}}]{1992ApJ...388...17W}
\bibinfo{author}{\bibfnamefont{A.~M.} \bibnamefont{{Wolfe}}},
  \bibinfo{author}{\bibfnamefont{K.~M.} \bibnamefont{{Lanzetta}}},
  \bibnamefont{and} \bibinfo{author}{\bibfnamefont{A.~L.}
  \bibnamefont{{Oren}}}, \bibinfo{journal}{\apj}
  \textbf{\bibinfo{volume}{388}}, \bibinfo{pages}{17} (\bibinfo{year}{1992}).

\bibitem[{\citenamefont{{Kronberg} et~al.}(1992)\citenamefont{{Kronberg},
  {Perry}, and {Zukowski}}}]{1992ApJ...387..528K}
\bibinfo{author}{\bibfnamefont{P.~P.} \bibnamefont{{Kronberg}}},
  \bibinfo{author}{\bibfnamefont{J.~J.} \bibnamefont{{Perry}}},
  \bibnamefont{and} \bibinfo{author}{\bibfnamefont{E.~L.~H.}
  \bibnamefont{{Zukowski}}}, \bibinfo{journal}{\apj}
  \textbf{\bibinfo{volume}{387}}, \bibinfo{pages}{528} (\bibinfo{year}{1992}).

\bibitem[{\citenamefont{Kojima et~al.}(2008, in press.)\citenamefont{Kojima,
  Yamazaki, and Kajino}}]{kojima}
\bibinfo{author}{\bibfnamefont{K.}~\bibnamefont{Kojima}},
  \bibinfo{author}{\bibfnamefont{D.~G.} \bibnamefont{Yamazaki}},
  \bibnamefont{and} \bibinfo{author}{\bibfnamefont{T.}~\bibnamefont{Kajino}},
  \bibinfo{journal}{AIP (American Institute of Physics) Conf. Proc. of OMEG07,
  Sapporo, Japan}  (\bibinfo{year}{2008, in press.}).

\bibitem[{\citenamefont{{Spergel} et~al.}(2007)\citenamefont{{Spergel}, {Bean},
  {Dor{\'e}}, {Nolta}, {Bennett}, {Dunkley}, {Hinshaw}, {Jarosik}, {Komatsu},
  {Page} et~al.}}]{2007ApJS..170..377S}
\bibinfo{author}{\bibfnamefont{D.~N.} \bibnamefont{{Spergel}}},
  \bibinfo{author}{\bibfnamefont{R.}~\bibnamefont{{Bean}}},
  \bibinfo{author}{\bibfnamefont{O.}~\bibnamefont{{Dor{\'e}}}},
  \bibinfo{author}{\bibfnamefont{M.~R.} \bibnamefont{{Nolta}}},
  \bibinfo{author}{\bibfnamefont{C.~L.} \bibnamefont{{Bennett}}},
  \bibinfo{author}{\bibfnamefont{J.}~\bibnamefont{{Dunkley}}},
  \bibinfo{author}{\bibfnamefont{G.}~\bibnamefont{{Hinshaw}}},
  \bibinfo{author}{\bibfnamefont{N.}~\bibnamefont{{Jarosik}}},
  \bibinfo{author}{\bibfnamefont{E.}~\bibnamefont{{Komatsu}}},
  \bibinfo{author}{\bibfnamefont{L.}~\bibnamefont{{Page}}},
  \bibnamefont{et~al.}, \bibinfo{journal}{ApJS} \textbf{\bibinfo{volume}{170}},
  \bibinfo{pages}{377} (\bibinfo{year}{2007}), \eprint{arXiv:astro-ph/0603449}.

\bibitem[{\citenamefont{{Ichikawa} et~al.}(2005)\citenamefont{{Ichikawa},
  {Fukugita}, and {Kawasaki}}}]{2005PhRvD..71d3001I}
\bibinfo{author}{\bibfnamefont{K.}~\bibnamefont{{Ichikawa}}},
  \bibinfo{author}{\bibfnamefont{M.}~\bibnamefont{{Fukugita}}},
  \bibnamefont{and}
  \bibinfo{author}{\bibfnamefont{M.}~\bibnamefont{{Kawasaki}}},
  \bibinfo{journal}{\prd} \textbf{\bibinfo{volume}{71}},
  \bibinfo{pages}{043001} (\bibinfo{year}{2005}),
  \eprint{arXiv:astro-ph/0409768}.

\bibitem[{\citenamefont{{Fukugita} et~al.}(2006)\citenamefont{{Fukugita},
  {Ichikawa}, {Kawasaki}, and {Lahav}}}]{2006PhRvD..74b7302F}
\bibinfo{author}{\bibfnamefont{M.}~\bibnamefont{{Fukugita}}},
  \bibinfo{author}{\bibfnamefont{K.}~\bibnamefont{{Ichikawa}}},
  \bibinfo{author}{\bibfnamefont{M.}~\bibnamefont{{Kawasaki}}},
  \bibnamefont{and} \bibinfo{author}{\bibfnamefont{O.}~\bibnamefont{{Lahav}}},
  \bibinfo{journal}{\prd} \textbf{\bibinfo{volume}{74}},
  \bibinfo{pages}{027302} (\bibinfo{year}{2006}),
  \eprint{arXiv:astro-ph/0605362}.

\bibitem[{\citenamefont{{Ma} and {Bertschinger}}(1995)}]{1995ApJ...455....7M}
\bibinfo{author}{\bibfnamefont{C.-P.} \bibnamefont{{Ma}}} \bibnamefont{and}
  \bibinfo{author}{\bibfnamefont{E.}~\bibnamefont{{Bertschinger}}},
  \bibinfo{journal}{\apj} \textbf{\bibinfo{volume}{455}}, \bibinfo{pages}{7}
  (\bibinfo{year}{1995}), \eprint{arXiv:astro-ph/9401007}.

\bibitem[{\citenamefont{{Yamazaki} et~al.}(2008)\citenamefont{{Yamazaki},
  {Ichiki}, {Kajino}, and {Mathews}}}]{2008arXiv0801.2572Y}
\bibinfo{author}{\bibfnamefont{D.~G.} \bibnamefont{{Yamazaki}}},
  \bibinfo{author}{\bibfnamefont{K.}~\bibnamefont{{Ichiki}}},
  \bibinfo{author}{\bibfnamefont{T.}~\bibnamefont{{Kajino}}}, \bibnamefont{and}
  \bibinfo{author}{\bibfnamefont{G.~J.} \bibnamefont{{Mathews}}},
  \bibinfo{journal}{ArXiv e-prints} \textbf{\bibinfo{volume}{801}}
  (\bibinfo{year}{2008}), \eprint{0801.2572}.

\bibitem[{\citenamefont{{Hu} and {White}}(1997)}]{1997PhRvD..56..596H}
\bibinfo{author}{\bibfnamefont{W.}~\bibnamefont{{Hu}}} \bibnamefont{and}
  \bibinfo{author}{\bibfnamefont{M.}~\bibnamefont{{White}}},
  \bibinfo{journal}{\prd} \textbf{\bibinfo{volume}{56}}, \bibinfo{pages}{596}
  (\bibinfo{year}{1997}), \eprint{arXiv:astro-ph/9702170}.

\bibitem[{\citenamefont{{Kasai} and {Tomita}}(1986)}]{1986PhRvD..33.1576K}
\bibinfo{author}{\bibfnamefont{M.}~\bibnamefont{{Kasai}}} \bibnamefont{and}
  \bibinfo{author}{\bibfnamefont{K.}~\bibnamefont{{Tomita}}},
  \bibinfo{journal}{\prd} \textbf{\bibinfo{volume}{33}}, \bibinfo{pages}{1576}
  (\bibinfo{year}{1986}).

\bibitem[{\citenamefont{{Durrer} et~al.}(2000)\citenamefont{{Durrer},
  {Ferreira}, and {Kahniashvili}}}]{2000PhRvD..61d3001D}
\bibinfo{author}{\bibfnamefont{R.}~\bibnamefont{{Durrer}}},
  \bibinfo{author}{\bibfnamefont{P.~G.} \bibnamefont{{Ferreira}}},
  \bibnamefont{and}
  \bibinfo{author}{\bibfnamefont{T.}~\bibnamefont{{Kahniashvili}}},
  \bibinfo{journal}{\prd} \textbf{\bibinfo{volume}{61}},
  \bibinfo{pages}{043001} (\bibinfo{year}{2000}),
  \eprint{arXiv:astro-ph/9911040}.

\bibitem[{\citenamefont{{Dodelson} et~al.}(1996)\citenamefont{{Dodelson},
  {Gates}, and {Stebbins}}}]{1996ApJ...467...10D}
\bibinfo{author}{\bibfnamefont{S.}~\bibnamefont{{Dodelson}}},
  \bibinfo{author}{\bibfnamefont{E.}~\bibnamefont{{Gates}}}, \bibnamefont{and}
  \bibinfo{author}{\bibfnamefont{A.}~\bibnamefont{{Stebbins}}},
  \bibinfo{journal}{\apj} \textbf{\bibinfo{volume}{467}}, \bibinfo{pages}{10}
  (\bibinfo{year}{1996}), \eprint{arXiv:astro-ph/9509147}.

\bibitem[{\citenamefont{{Lesgourgues} and
  {Pastor}}(2006)}]{2006PhR...429..307L}
\bibinfo{author}{\bibfnamefont{J.}~\bibnamefont{{Lesgourgues}}}
  \bibnamefont{and} \bibinfo{author}{\bibfnamefont{S.}~\bibnamefont{{Pastor}}},
  \bibinfo{journal}{Phys. Rept.} \textbf{\bibinfo{volume}{429}},
  \bibinfo{pages}{307} (\bibinfo{year}{2006}), \eprint{arXiv:astro-ph/0603494}.

\bibitem[{\citenamefont{Lewis et~al.}(2000)\citenamefont{Lewis, Challinor, and
  Lasenby}}]{Lewis:1999bs}
\bibinfo{author}{\bibfnamefont{A.}~\bibnamefont{Lewis}},
  \bibinfo{author}{\bibfnamefont{A.}~\bibnamefont{Challinor}},
  \bibnamefont{and} \bibinfo{author}{\bibfnamefont{A.}~\bibnamefont{Lasenby}},
  \bibinfo{journal}{Astrophys. J.} \textbf{\bibinfo{volume}{538}},
  \bibinfo{pages}{473} (\bibinfo{year}{2000}), \eprint{astro-ph/9911177}.

\bibitem[{\citenamefont{{Caprini} and {Durrer}}(2002)}]{2002PhRvD..65b3517C}
\bibinfo{author}{\bibfnamefont{C.}~\bibnamefont{{Caprini}}} \bibnamefont{and}
  \bibinfo{author}{\bibfnamefont{R.}~\bibnamefont{{Durrer}}},
  \bibinfo{journal}{\prd} \textbf{\bibinfo{volume}{65}},
  \bibinfo{pages}{023517} (\bibinfo{year}{2002}),
  \eprint{arXiv:astro-ph/0106244}.

\bibitem[{\citenamefont{{Kamionkowski}
  et~al.}(1997)\citenamefont{{Kamionkowski}, {Kosowsky}, and
  {Stebbins}}}]{1997PhRvD..55.7368K}
\bibinfo{author}{\bibfnamefont{M.}~\bibnamefont{{Kamionkowski}}},
  \bibinfo{author}{\bibfnamefont{A.}~\bibnamefont{{Kosowsky}}},
  \bibnamefont{and}
  \bibinfo{author}{\bibfnamefont{A.}~\bibnamefont{{Stebbins}}},
  \bibinfo{journal}{\prd} \textbf{\bibinfo{volume}{55}}, \bibinfo{pages}{7368}
  (\bibinfo{year}{1997}), \eprint{arXiv:astro-ph/9611125}.

\end{thebibliography}
\end{document}